\documentclass[10pt]{article}

\usepackage{amsmath, amsthm, amscd, amssymb, graphicx, float}

\newcommand{\vect}[1]{\textbf{\textit{#1}}} 
\newcommand{\bb}[1]{\mathbb{#1}}            
\renewcommand{\Im}{\mathfrak{Im}}           

\title{Quantum Dynamics with Bohmian Trajectories}

\author{D.-A. Deckert\thanks{Mathematisches Institut,
        LMU M\"unchen, dirk.deckert@mathematik.uni-muenchen.de},
        D. D\"urr\thanks{Mathematisches Institut,
        LMU M\"unchen, duerr@mathematik.uni-muenchen.de},
        P. Pickl\thanks{Institut f\"ur theoretische Physik,
        Universit\"at Wien, pickl@mathematik.uni-muenchen.de}}

\date{January 15, 2007}

\begin{document}

\maketitle

\begin{abstract}
We describe the advantages and disadvantages of numerical methods
when Bohmian trajectory-grids are used for numerical simulations of quantum dynamics. We focus on the crucial non-crossing property of
Bohmian trajectories, which numerically must be paid careful
attention to. Failure to do so causes instabilities or leads to
false simulations.
\end{abstract}


\section{Introduction}
\paragraph{Bohmian Mechanics.} Bohmian Mechanics \cite{BM01} is a
mechanical theory for the motion of particles. We describe the
theory in units of mass $m=\hbar=1$. Given the Schr\"odinger wave
function $\psi_t$ of an $N$-particle system, the trajectories of
the $N$ particles $\vect q_k(\vect q_k^0,\psi^0;t)\in\bb R^3$,
$1\leq k\leq N$, are solutions of
\begin{equation}\label{Bohm}
  \frac{d\vect q_k(t)}{dt} = \Im\bigg[
    \frac{\psi^*_t \nabla_{\vect q_k} \psi_t}{\psi^*_t \psi_t}(\vect q_1,...,\vect q_N)\bigg]
\end{equation}
with $\vect q_k(t=0)=\vect q_k^0$ as initial conditions and
$\psi^*$ denoting the complex conjugate of $\psi$, so that
$\psi^*\psi=|\psi|^2$.\footnote{Interpreting $\psi^*\psi$ as inner
product, the generalization to spin wave functions is straight
forward.} We denote the configuration point $x=(\vect
x_1,...,\vect x_N)\in\bb R^{3N}$, $\nabla_x=(\nabla_{\vect
x_1},\ldots,\nabla_{\vect x_N})$ 
where
$\nabla_{\vect x_k}$ is the
gradient with respect to $\vect x_k\in\bb R^3$. $\psi_t$ is the
solution of the Schr\"odinger equation
\begin{align}\label{eqn:schroedinger}
  i\frac{d \psi_t(x)}{d t} &= \bigg(-\frac{\nabla_x^2}{2}+V(x)\bigg)
    \psi_t(x)
\end{align}
with the initial condition $\psi_{t=0}(x)=\psi_{t=0}(\vect x_1,...,\vect
x_N)=\psi^0(\vect x_1,...,\vect x_N)$. Using configuration space
language the Bohmian trajectory of an $N$ particle system is an
integral curve $q(t)\in\bb R^{3N}$ of the following velocity field
on configuration space
\begin{align}\label{eqn:velField}
  v^{\psi}(q, t) = \Im\bigg[\frac{(\psi_t^*\nabla_q\psi_t)(q)} {(\psi_t^*\psi_t)(q)}\bigg]
\end{align}
i.e.
\begin{align}\label{eqn:configBohm}
  \frac{d q(t)}{dt}= v^{\psi}(q,t)
\end{align}
Under general conditions one has global existence and uniqueness of Bohmian
trajectories, i.e. the integral curves do not run into nodes of the wave functions  and they cannot cross
\cite{BMExistenceProof05}. From now on we shall only talk about
the trajectories as integral curves in configuration space. Note
that for one particle the configuration space is equal to physical
space, for more than one particle
this is not the case. \\

The empirical import of   Bohmian Mechanics arises from
equivariance of the $|\psi|^2$-measure:  One readily sees that by
virtue of (\ref{eqn:velField}) the continuity equation for the
Bohmian flow on configuration space is identically fulfilled  by
the density $\rho_t=|\psi_t|^2$,  then known as the quantum flux
equation,
\begin{align*}
  \frac{\partial |\psi(x,t)|^2}{\partial t} + \nabla_x \left(
|\psi(x,t)|^2v^\psi(x,t)\right)=0
\end{align*} This means that if
the configuration of particles $q=(\vect q^0_1,$ $...,\vect
q^0_N)$ is distributed according to $\rho_0=|\psi^0(\vect
x_1,...\vect x_N)|^2$ at time $t=0$, then the configuration
$q(t)=(\vect q_1(\vect q_1^0,\psi^0;t),$ $...,\vect q_N(\vect
q_N^0,\psi^0;t))$ is distributed according to  $\rho_t=
|\psi_t(\vect x_1,...\vect x_N)|^2$ at any time $t$. Because of
this, Bohmian Mechanics agrees with all predictions made by
orthodox quantum mechanics whenever the latter are unambiguous \cite{QEpaper, OPpaper}.\\
\paragraph{Hydrodynamic Formulation of Bohmian Mechanics.} Write $\psi$  in the Euler form
\begin{align*} \psi(x,t)=R(x,t)e^{iS(x,t)}
\end{align*} where $R$ and $S$ are given by real-valued functions.
Equation (\ref{eqn:velField}) together with equation
(\ref{eqn:schroedinger}) separated in 
their
real and complex parts
gives the following set of differential equations \begin{align*}
\frac{d q}{dt} &= \nabla_{q}S(q,t)\\ \frac{d R(x,t)}{dt} &=
-\frac{1}{2}\nabla_{x}(R(x,t)\nabla_{x}S(x,t))\\ \frac{d
S(x,t)}{dt} &= -\frac{1}{2}\left(\nabla_x
S(x,t)\right)^2-V(r)+\frac{1}{2}\frac{\nabla_{x}^2 R(x,t)}{R(x,t)}
\end{align*} This set of equations can be examined along the
Bohmian trajectory
$q$. We then obtain
\begin{align}\label{eqn:diffEqs1}
  \frac{d q}{dt} &= \nabla_{q}S(q,t)\\
  \frac{\partial R(q,t)}{\partial t} &= -\frac{1}{2}R(q,t)\nabla_{q}^2S(q,t)\label{eqn:diffEqs2}\\
  \frac{\partial S(q,t)}{\partial t} &= \frac{1}{2}\left(\frac{dq}{dt}\right)^2-V(q)+\frac{1}{2}
  \frac{\nabla_{q}^2 R(q,t)}{R(q,t)}\label{eqn:diffEqs3}
\end{align}

\paragraph{Numerical Integration.} Numerically this set of
equations (\ref{eqn:diffEqs1})-(\ref{eqn:diffEqs3}) can readily be
integrated \cite{Wyatt} and offers some advantages over standard
techniques for solving the Schr\"odinger equation
(\ref{eqn:schroedinger}) numerically. The basic idea, due to Wyatt
\cite{Wyatt}, is that the Bohmian configuration space trajectories
define a co-moving grid (Bohmian grid) in configuration space
which is best adapted for the computation of $\psi$. If initially
the grid points ($n$ points in $\bb R^{3N}$) are  $|\psi^0|^2$
distributed, they will remain $|\psi_t|^2$ distributed for all
times $t$. So the co-moving grid spreads dynamically according to
the spreading of $|\psi_t|^2$ and the grid points will primarily
remain in regions of space where $|\psi_t|^2$ is large while
avoiding regions of nodes or tails of $|\psi_t|^2$ which are
numerically problematic. Therefore Wyatt's idea is this: Integrate
the equations (\ref{eqn:diffEqs1})-(\ref{eqn:diffEqs3}) {\em
simultaneously, i.e. get the best adapted moving grid along with
$\psi$}, instead of integrating the Schr\"odinger equation on some
fixed or otherwise determined grid. This is why the algorithm is
particularly interesting for long-time simulations which usually
demand huge computational effort on fixed grids. Therefore if one
aims at the relevant parts of the wave function (where
probabilities are high) one can use a fixed number $n$ of
$|\psi|^2$ distributed Bohmian grid points in $\bb R^{3N}$, so
that the Bohmian grid simulation scales with the number $N$ of
particles \cite{Wyatt} while conventional grid methods mostly
scale exponentially with $N$. We say more in section \ref{dist}
on how to distribute $n$ grid points in a $|\psi|^2 $ manner.\\

In order to perform the numerical integration of the set of
differential equations (\ref{eqn:diffEqs1})-(\ref{eqn:diffEqs3})
we follow the straight-forward method described in \cite{Wyatt}.
The only crucial part in this is computing the derivatives
involved in the set of differential equations. There are several
techniques known  and \cite{Wyatt} gives a comprehensive overview.
Among them one technique called \emph{least square fitting} is
commonly used. In this letter we argue that this method is
inappropriate for integrating
(\ref{eqn:diffEqs1})-(\ref{eqn:diffEqs3}) for general initial
conditions and potentials. Bad situations arise whenever Bohmian
trajectories move towards each other, because least square fitting
will allow crossings of the simulated trajectories which are not
allowed for Bohmian trajectories. Once a crossing is encountered
in a numerical simulation further computation can be aborted
because this numerical wave function would differ immensely from
the solution of the Schr\"odinger equation
(\ref{eqn:schroedinger}). This will happen generically, i.e. for
non-gaussian wave functions. To illustrate our argument we shall
present a typical numerical example in one dimension with one
particle.
Therefore in what follows $N =1$ and the $n$ trajectories which we
shall consider (making up the grid) are the possible trajectories
of this
one
particle only.
\section{Least Square versus Polynomial Fitting}\label{sec:fitting}

In order to compute the derivatives encountered in
(\ref{eqn:diffEqs1})-(\ref{eqn:diffEqs3}) we only consider two
different types of fitting, the \emph{least square fitting} and
the \emph{polynomial fitting}. Most other fitting algorithms are
descendants of one or the other. Both provide an algorithm for
finding e.g. a polynomial\footnote{Or more general an element of a
$m$ dimensional vector space of functions. The coefficients of the
design matrix $X$ determine which basis functions are used.} of degree,
say $(m-1)$, which is in some sense to be specified close to a
given function $f:\bb R\to\bb R$ known only on a set of, say $n$,
pairwise distinct data points, $(x_i,f(x_i))_{1\leq i\leq n}$, as
subset of the graph of $f$. The derivative can then be computed
from the fitting polynomial by algebraic means. Let for the
further discussion
\begin{align*}
  y      &:= (f(x_i))_{1\leq i\leq n}\\
  X      &:= (x_i^{(j-1)})_{1\leq i\leq n, 1\leq j\leq m}\\
  a      &=  (a_i)_{1\leq i\leq m}\in\bb R^m\\
  \delta &=  (\delta_i)_{1\leq i\leq n}\in\bb R^n
\end{align*}
Using this notation the problem of finding a fitting polynomial to $f$ on the
basis of $n$ pairwise distinct data points of the graph of $f$
reduces to finding the coefficients of the vector $a$ obeying the
equation
\begin{align*}
  y = X\cdot a+\delta
\end{align*}
such that the error term $\delta$ is in some sense small. Here the
dot
$\cdot$ denotes matrix multiplication. The fitting polynomial is
given by $p(x)=\sum_{j=1}^{m}a_j x^{j-1}$.

\paragraph{Polynomial Fitting.} 
For the case $n=m$ we can choose the error term $\delta$ to be identical zero since $X$ is 
an
invertible square matrix and $a=X^{-1}\cdot y$ can be straight-forwardly computed.

\paragraph{Least Square Fitting.} 
For arbitrary $n\geq m$ there is no unique solution anymore and one needs a new criterion for finding a unique vector $a$. The algorithm of least square fitting uses therefore the minimum value of the accumulated error $\Delta:=\sum_{i=1}^n w(x-x_i)\delta_i^2$ where $w:\bb R\to\bb R$ specifies a weight dependent on the distance between the $i$-th data point $x_i$ and some point $x$ where the fitting polynomial shall be evaluated. The vector $a$ can now be determined by minimizing $\Delta$ as a function of $a$ by solving
\begin{align*}
  \left(\frac{\partial \Delta(a)}{\partial a_j}\right)_{1\leq j\leq m} = \left(-2\sum_{i=1}^n w(x-x_i)(y-X\cdot a)_i x_i^{j-1}\right)_{1\leq j\leq m}=0
\end{align*}
Note that for $m=n$ and any non-zero weight $w$ the algorithm of
least square fitting produces the same $a$ as the polynomial
fitting would do since for the $a$ determined by polynomial
fitting the non-negative function $\Delta(a)$ is zero and thus $a$
naturally minimizes the error term.

\paragraph{Why Least Square Fitting is inappropriate for Bohmian
Grids.}
The algorithm of least square fitting is well known for
its tendency to stabilize numerical simulations by averaging out
numerical errors. But 
exactly
this averaging makes it hard to keep the
grid points from crossing each other.   In order to understand
what happens during a numerical simulation recall the form of the
equations of motion (\ref{eqn:diffEqs1})-(\ref{eqn:diffEqs3})
which have to be integrated step by step. The change in time of
the phase $S$ is determined by three terms in equation of
(\ref{eqn:diffEqs3}). The first two terms form the classical
Lagrangian and depend on the current velocities of the grid points
and on the potential $V$. The third term is the so-called quantum
potential and depends only on $R$ and its second derivative. The
quantum potential is what prevents Bohmian trajectories from
crossing each other. It therefore needs special attention during
the numerical integration. Now imagine initial conditions such
that two grid points approach each other. An increase of the
density of the grid points in a region where the two grid points
move towards each other will cause a small bump in $R$.
By small bump we mean a bump  of the wave function shape on a
microscopic scale, i.e. 
a
scale defined  by few grid points. Such a
small bump of $R$ may have large derivatives changing the quantum
potential in (\ref{eqn:diffEqs3}).
 Thus, it is absolutely vital for
a numerical simulation to implement a fitting algorithm that
reconstructs not only $R$, respectively $S$, in an accurate way
but also its second derivative. The tendency of the least square
fitting algorithm is to average out those small bumps in $R$,
respectively $S$. Hence, the simulation is blind to recognize grid
points moving towards each other and hence does not prevent a
crossing of these trajectories.

Note that since the averaging occurs on a microscopic scale,
situations in which grid points move only very slowly or do
generically not move towards each other\footnote{For example a free
Gaussian wave packet or one approaching a potential creating only
soft reflections.} are often numerically doable with least square
fitting. On the other hand numerical simulations with least square
fitting in general situations like the one discussed above are bound
to break down as soon as two grid points get too close to each
other.

\paragraph{Why Polynomial Fitting is more appropriate for Bohmian Grids.}
We stressed above that small bumps may exponentiate numerical
instability. A good numerical method must take note of the small
bumps and must prevent their increase. Such a numerical method is
provided by polynomial fitting, since there the polynomials go
through all grid points. Therefore polynomial fitting recognizes
the bumps of $R$ and/or $S$ and hence the resulting quantum
potential recognizes the approaching grid points. Now recall the
physics of the Bohmian evolution,  which as we stressed in the
introduction prevents trajectories from crossing each other.
Therefore we expect that this method is self-correcting and hence
stabilizing.

\begin{figure}[hb]
\centering
\begin{tabular}{cc}
  \includegraphics[scale=0.25]{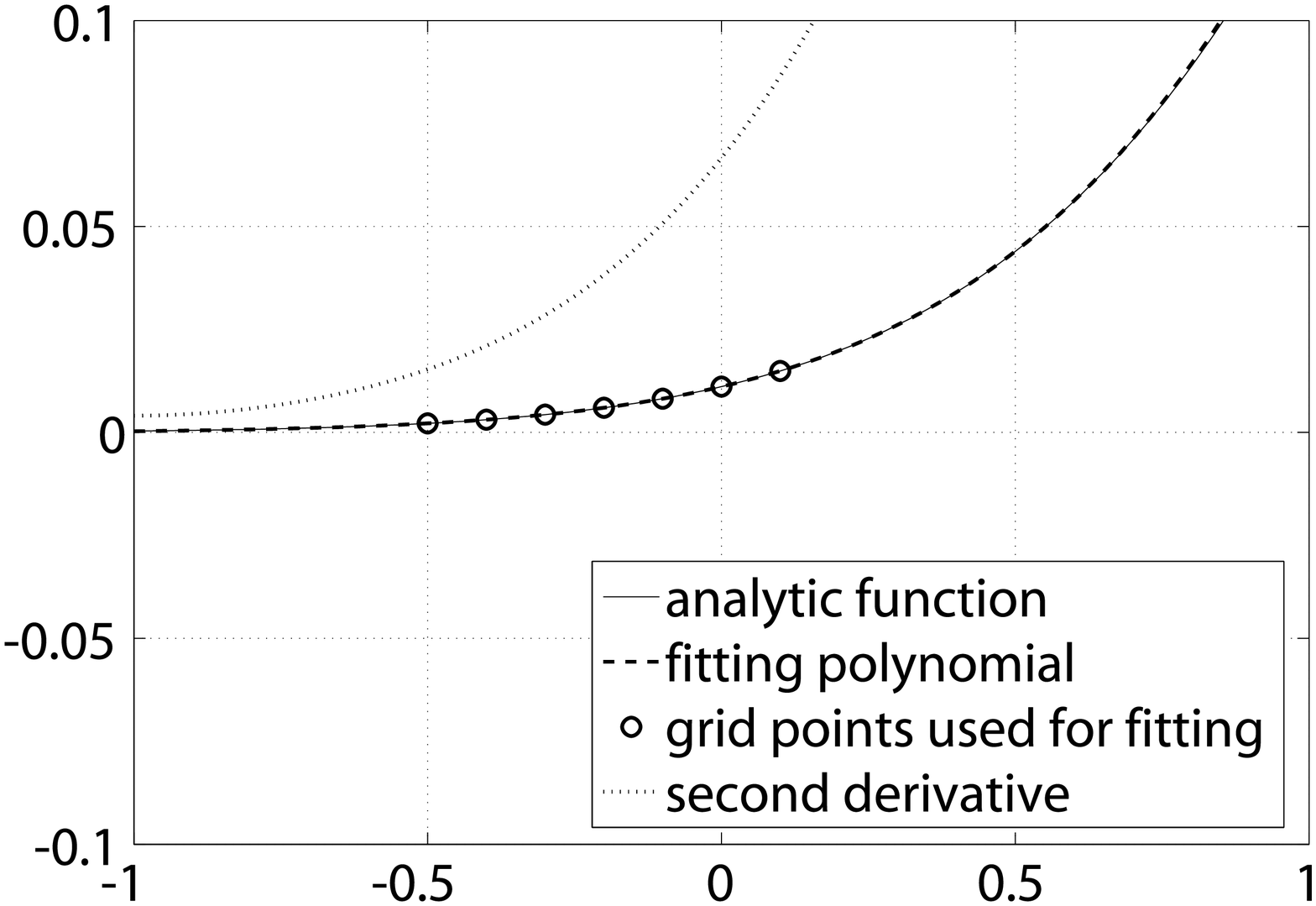} & \includegraphics[scale=0.25]{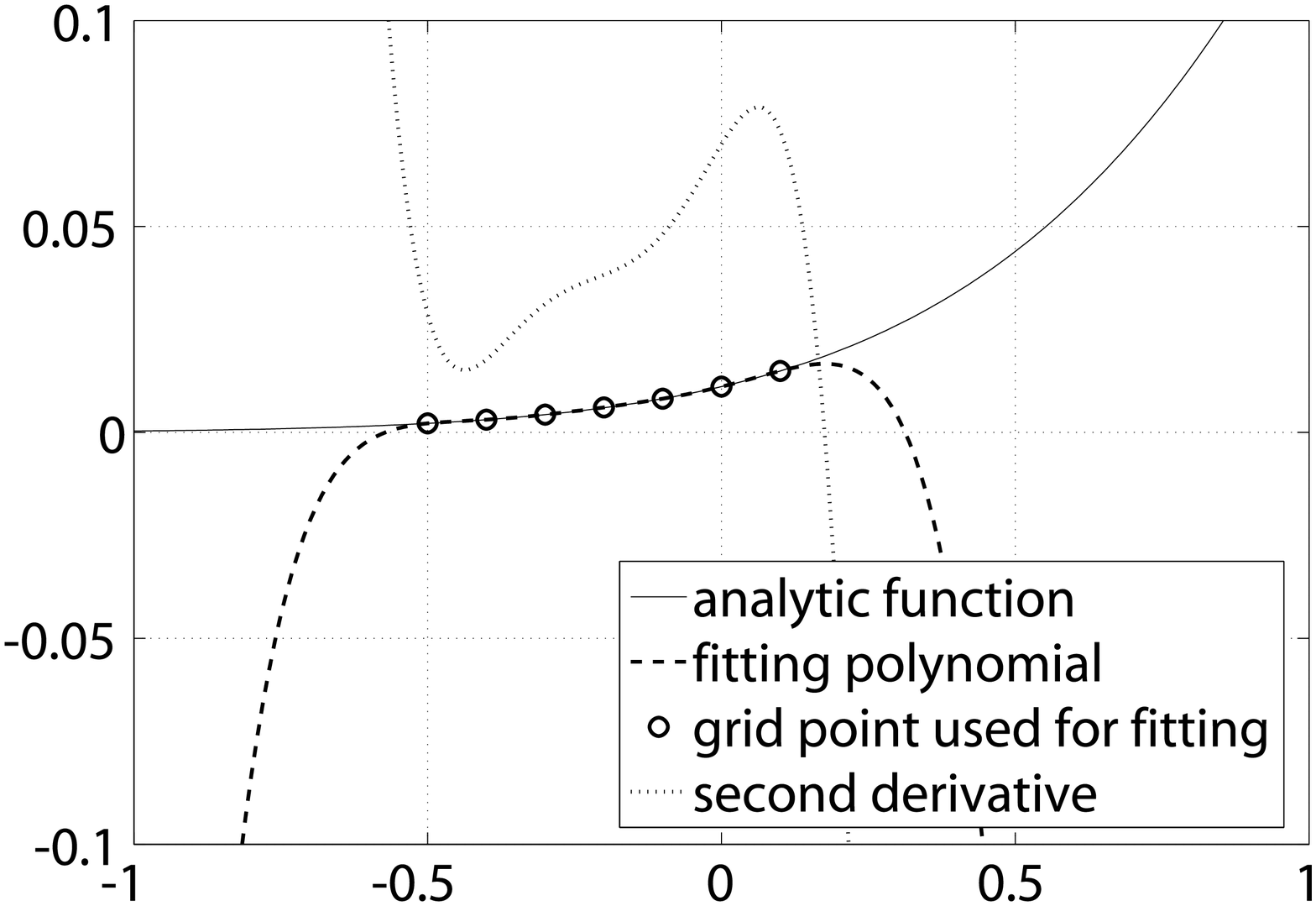}\\
  \includegraphics[scale=0.25]{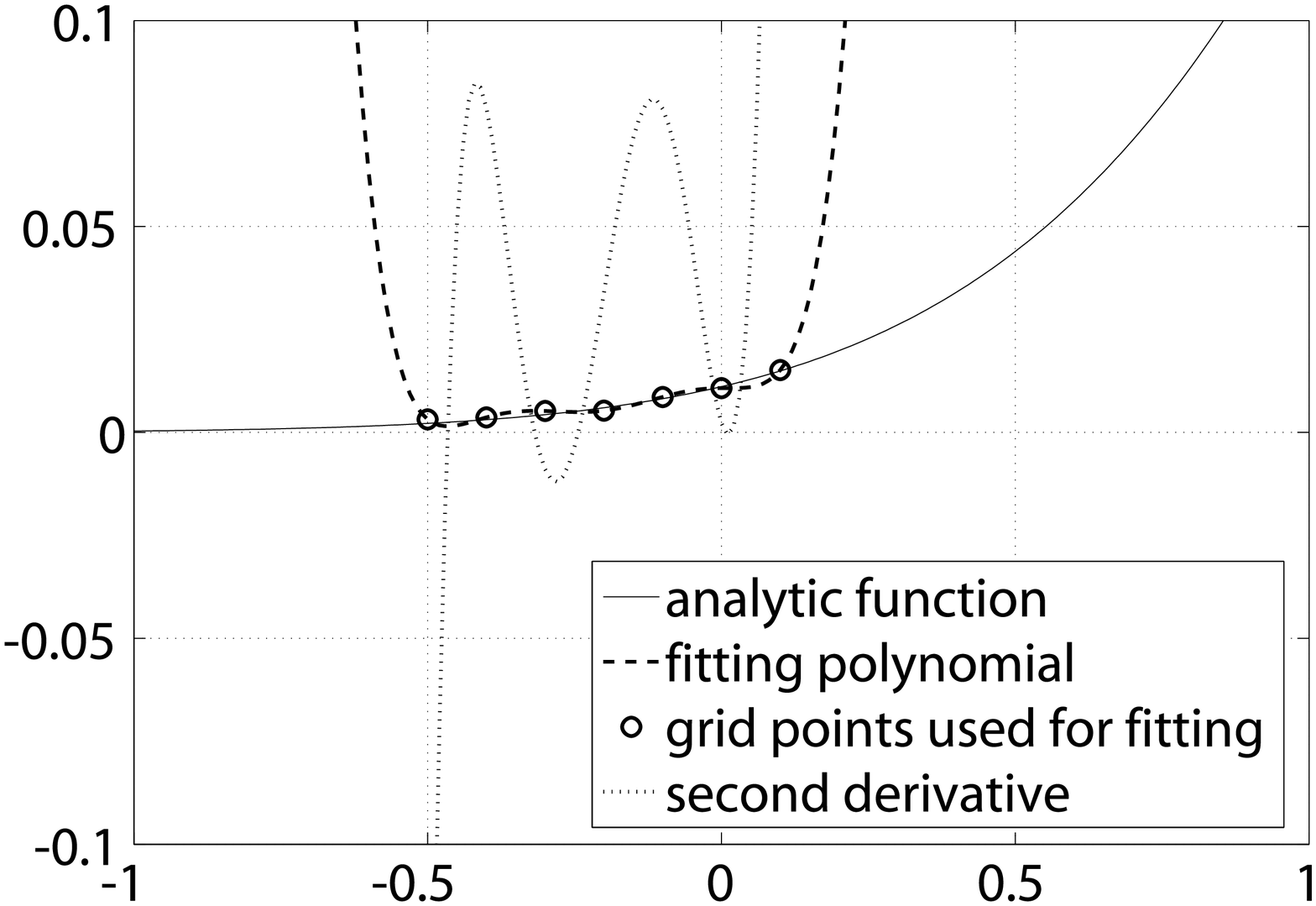} & \includegraphics[scale=0.25]{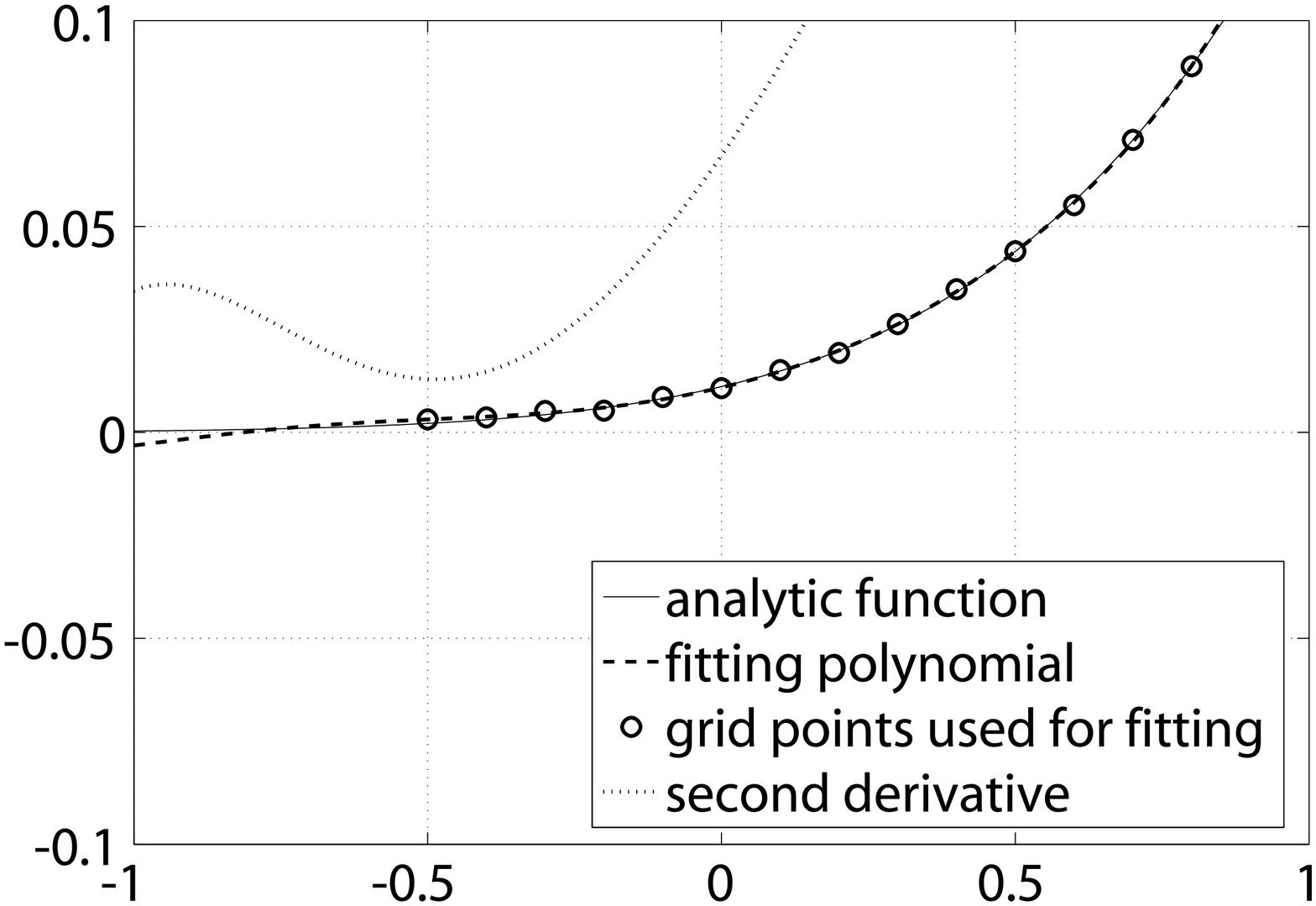}
\end{tabular}
\caption{Fitting polynomial of sixth degree and its second
derivative near the boundary. Upper left: the function to fit
smoothly decays. Both least square and polynomial fitting give
identical fitting polynomials mimicking this decay. Upper right:
polynomial fitting was used while the function value at fourth
grid point was shifted upwards by $10^{-4}$ (not visible in plot)
to simulate a numerical error in order to see how sensitive polynomial fitting reacts to such an error. Lower left: all values at the grid
points were shifted by an individual random amount of the order of
$10^{-3}$ and polynomial fitting was used. Lower right: all values
at the grid points were shifted by the same random numbers as on
the lower left while the additional grid points were also shifted
by individual random amounts of the order of $10^{-3}$ and least
square fitting was used. By comparison with upper left plot one observes the robustness of least square fitting to such numerical errors 
at the boundary. 
} \label{fig:boundary}
\end{figure}

\paragraph{The Boundary Problem.} It has to be remarked that
polynomial fitting creates a more severe problem at the boundary
of the supporting grid than least square fitting. This problem is
of conceptual kind since at the boundary there is a generic lack
of knowledge of how the derivatives of $R$ or $S$ behave, see
figure \ref{fig:boundary}.

We describe briefly how this conceptual
problem can cause severe numerical instability. For this
suppose that only the last grid point in the upper left plot of
figure \ref{fig:boundary} is lifted a little bit upwards by e.g.
some numerical error. Then the resulting quantum potential will
cause the last grid point to move towards the second last one.
This increases the density of grid points and thus again increases
$R$ in the next step such that this effect is self-amplifying (in fact growing
super exponentially) and will effect the whole wave function
quickly.

At the boundary  the good property of polynomial fitting, namely
to recognize all small bumps works against Bohmian grids
techniques. In contrary least square fitting simply averages these
small numerical errors out, see lower right plot in figure
\ref{fig:boundary}. The conceptual boundary problem, however,
remains and will show up eventually also 
with least square fitting.

\begin{figure}[hb]
\centering
\begin{tabular}{cc}
  \includegraphics[scale=0.24]{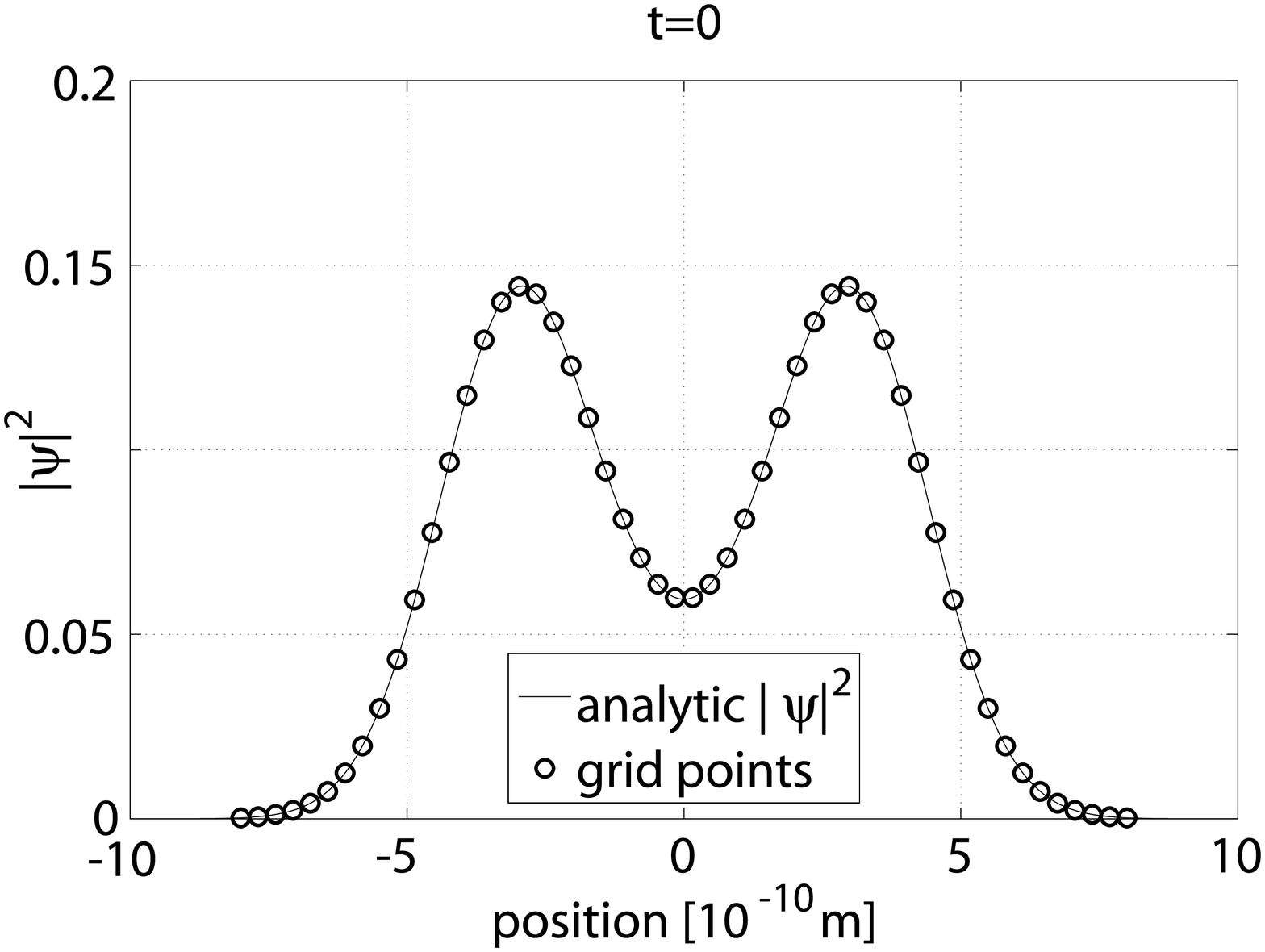} & \includegraphics[scale=0.24]{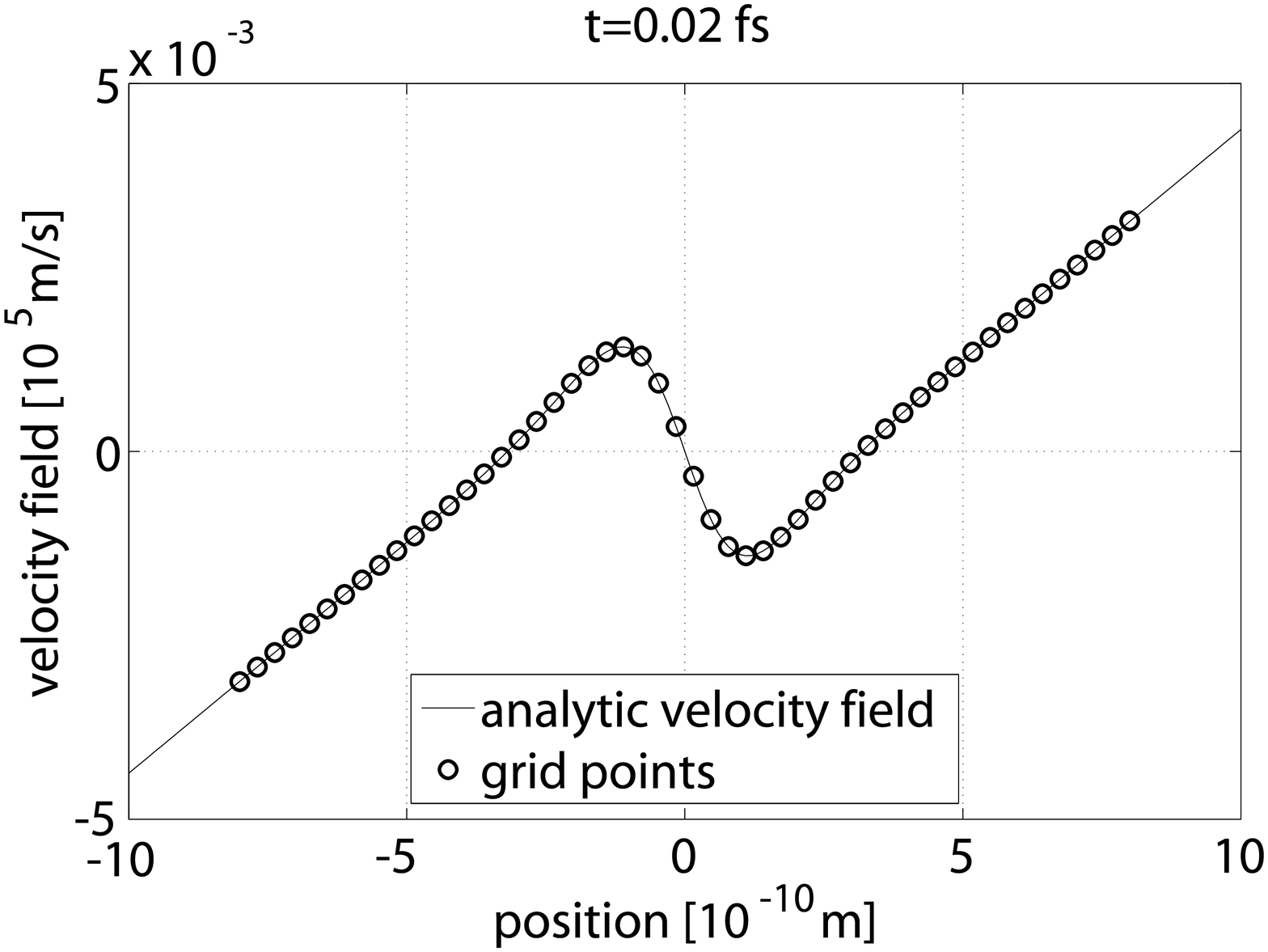}
\end{tabular}
\caption{Left: Initial wave function at time zero. Right: Its
velocity field after a very short time.}
\label{fig:initialconditions}
\end{figure}

\paragraph{The Numerical Simulation.} To demonstrate our
argument we shall now give a numerical example for one particle in
one dimension. We remark as mentioned earlier that Gaussian wave packets are unfit for probing the quality of the numerical simulations. Therefore we
take as initial conditions two superposed free Gaussian wave
packets with a small displacement together with a velocity field
identically to zero and focus the attention on the region where
their tails meet, i.e. where the Bohmian trajectories
will
move towards
each other according to the spreading of the wave packets, see
figure \ref{fig:initialconditions}. The physical units given in the figures and the following discussion refer to a wave packet on a length scale of $1\text{\AA} =10^{-10}\text{m}$ and a Bohmian particle with the mass of an electron.

The numerical simulation is implemented as in \cite{Wyatt} with minor changes. It is written in MATLAB using IEEE Standard 754 double precision. In order to ease the boundary problem we use a very strong least square fitting at the boundary. Note that this only eases the boundary problem and is by no means a proper cure. In between the boundary, however, the number of data points for the fitting algorithm and the degree of the fitting polynomial can be chosen in each run. In this way it is possible to have either polynomial fitting or least square fitting in between the boundary. The simulation is run twice. First with the polynomial fitting and then with the least square fitting algorithm in between the boundaries.

We take $7$ basis functions for the fitting polynomial in both cases. Note that
the choice should at least be greater or equal to $4$ in order to
have enough information about the third derivative of the fitting
polynomial. In the first run with polynomial fitting the number of
grid points used for fitting is $7$ and in the second run for the
least square fitting we choose $9$ which induces a mild least
square behavior. For the numerical integration we have used a time
step of $10^{-2}\text{fs}$ with the total number of $51$ grid points
supporting the initial wave function. Please find the source code of our numerical simulation at the end of this letter.

\begin{figure}[hb]
\centering
\begin{tabular}{cc}
  \includegraphics[scale=0.24]{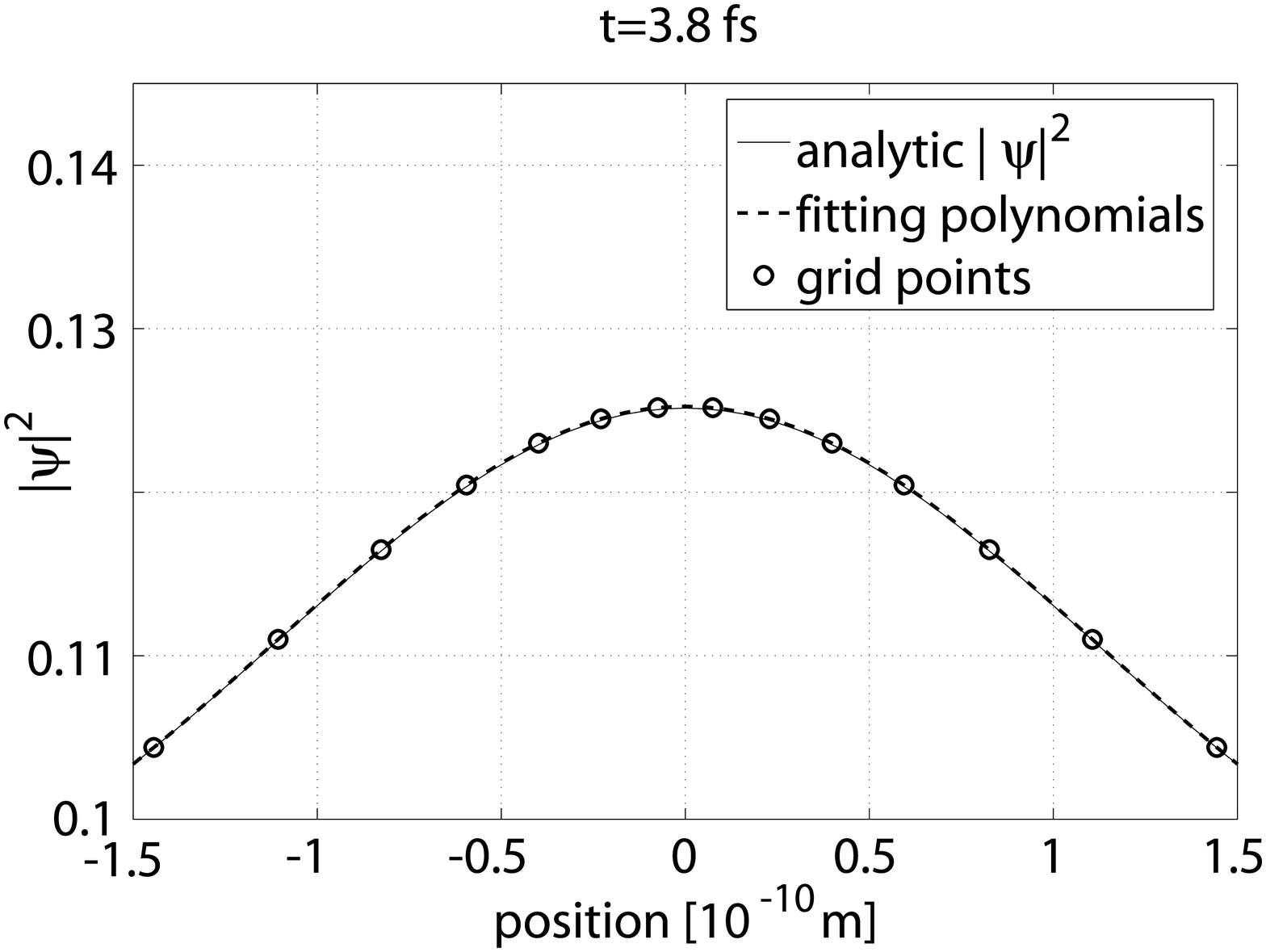} & \includegraphics[scale=0.24]{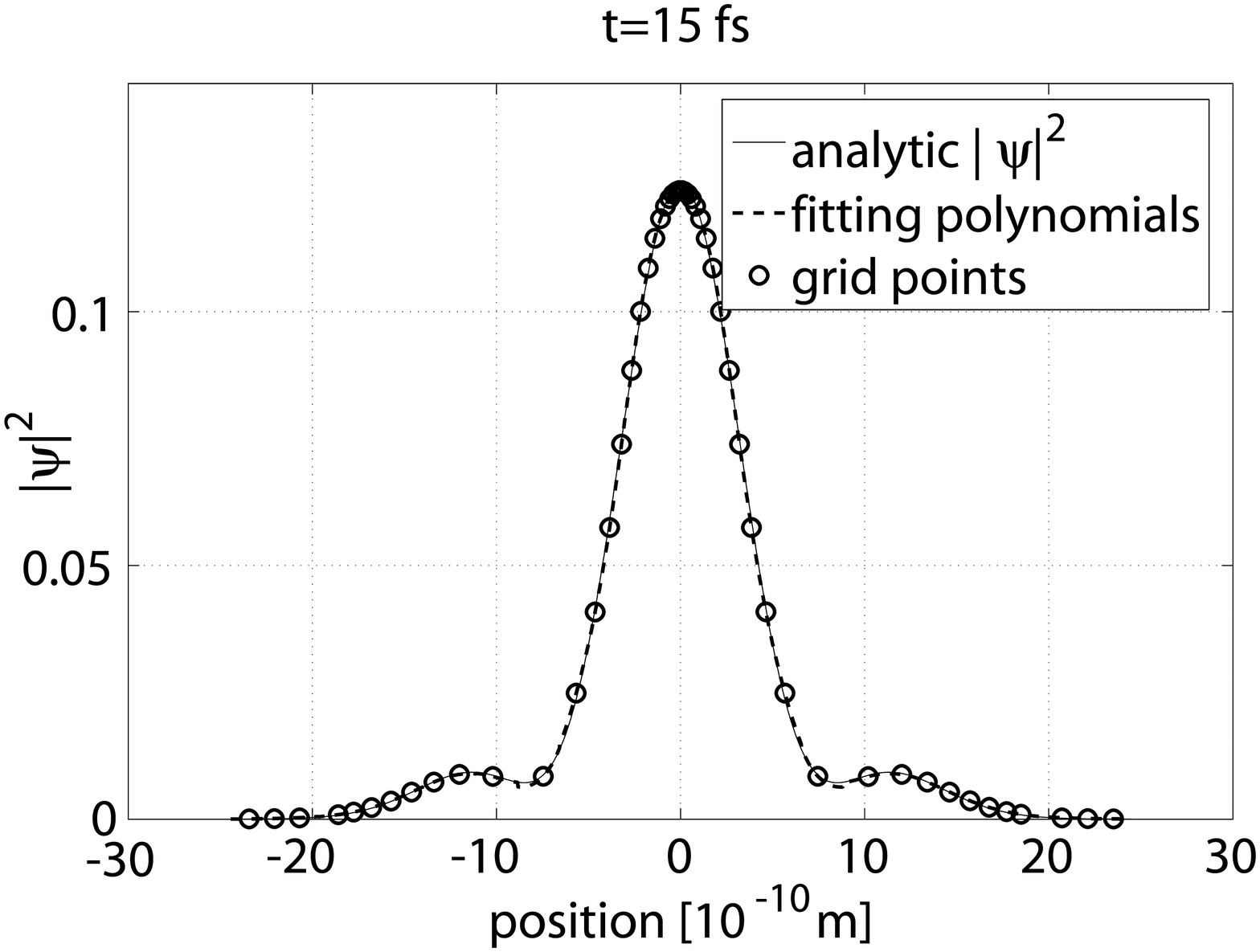}\\
  \includegraphics[scale=0.24]{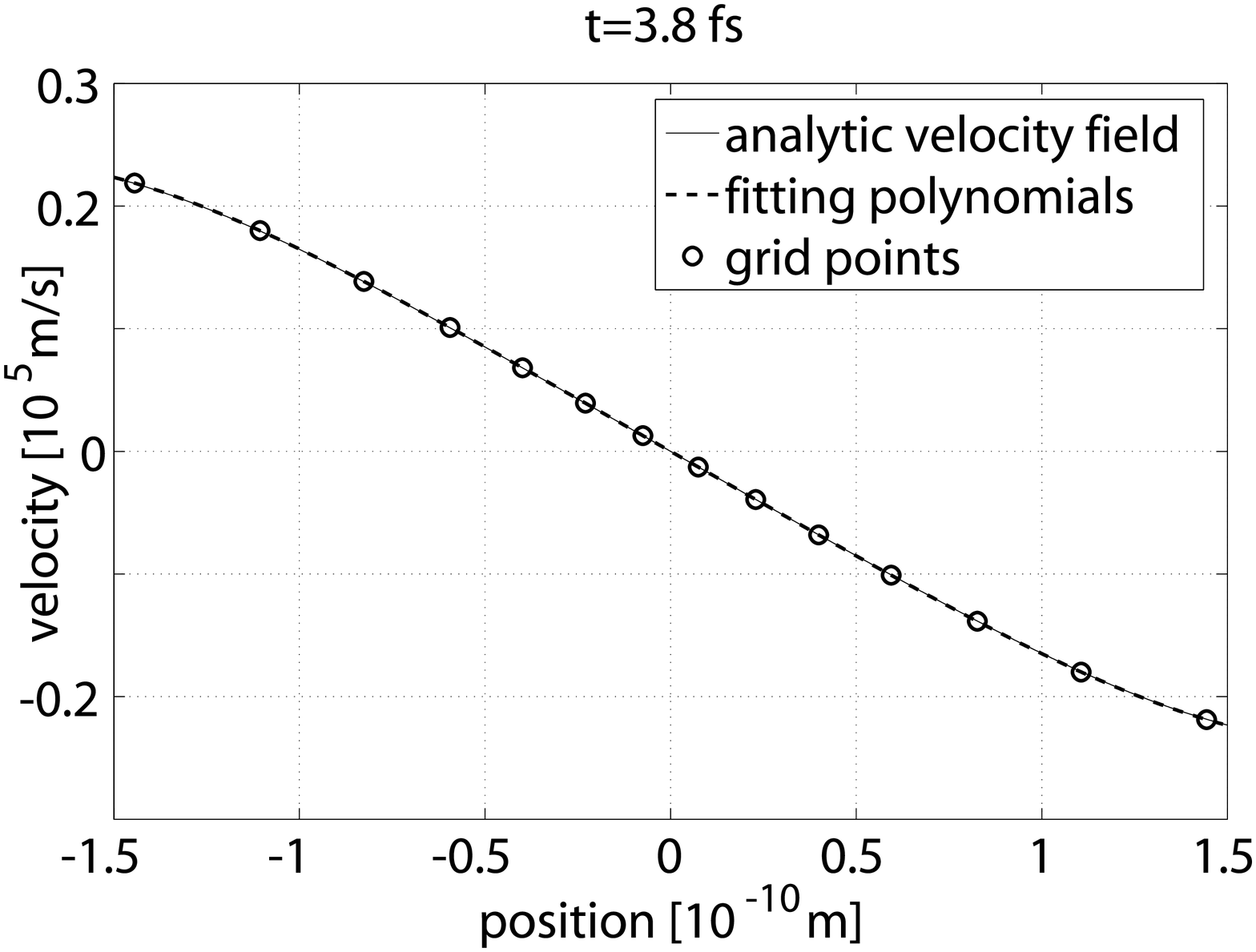} & \includegraphics[scale=0.24]{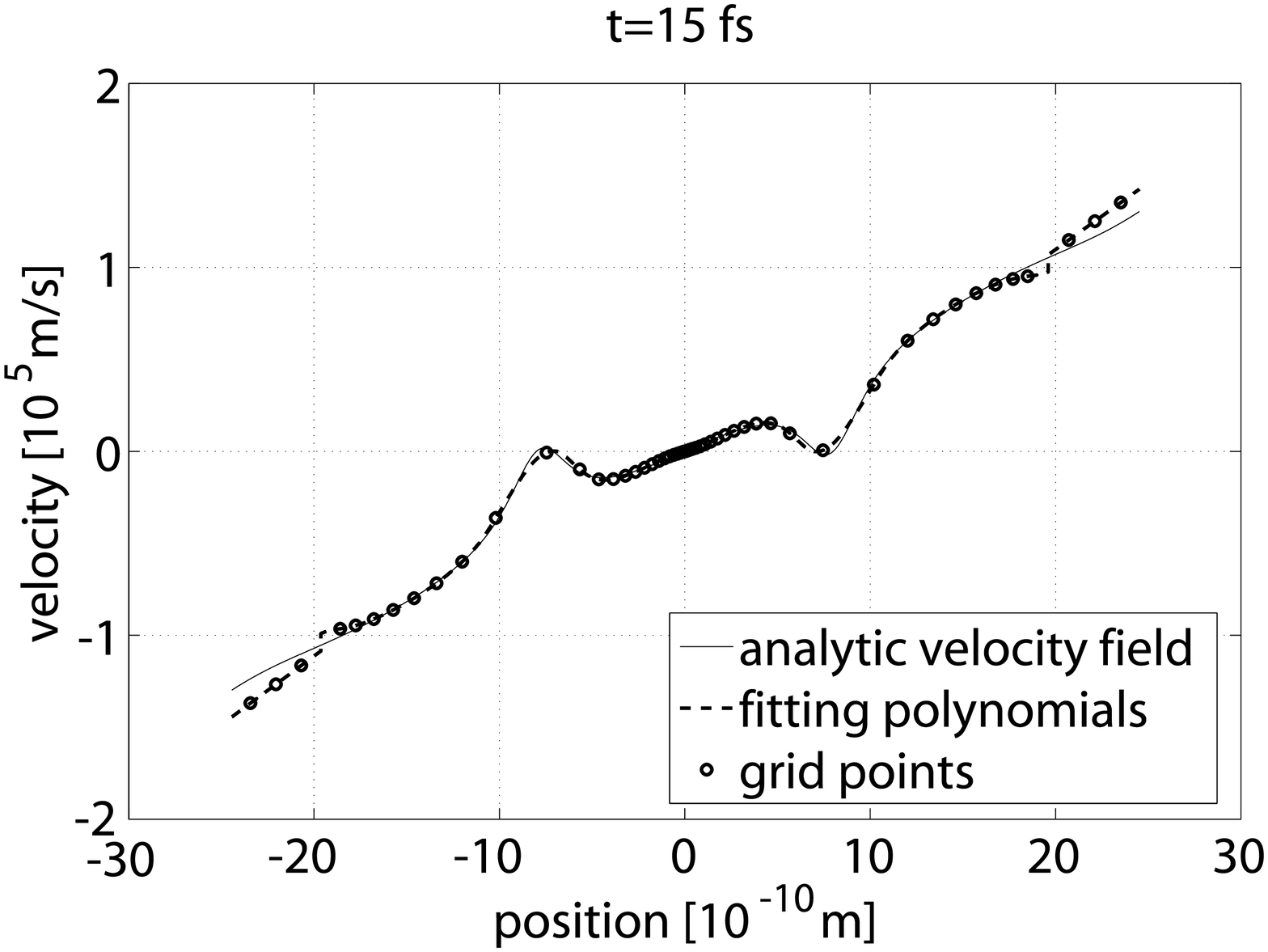}\\
\end{tabular}
\caption{Simulated $|\psi|^2$ together with the velocity field at
times $t=3.8\text{fs}$ and $t=15\text{fs}$ using polynomial fitting in
between the boundary. Left: the center part of the wave function
where grid points move towards each other. Right: the whole wave
function at a later time. The grid points have all turned and move
apart from each other. 
The kinks at $\pm 20\cdot 10^{-15}m$ in the velocity field in the
lower right plot are due to the transition from least square fitting at the boundary to polynomial fitting in between the boundary (see paragraph: The Boundary Problem in section \ref{sec:fitting}).
} \label{fig:simulation_polyfit}
\end{figure}

\begin{figure}[hb]
\centering
\begin{tabular}{cc}
  \includegraphics[scale=0.24]{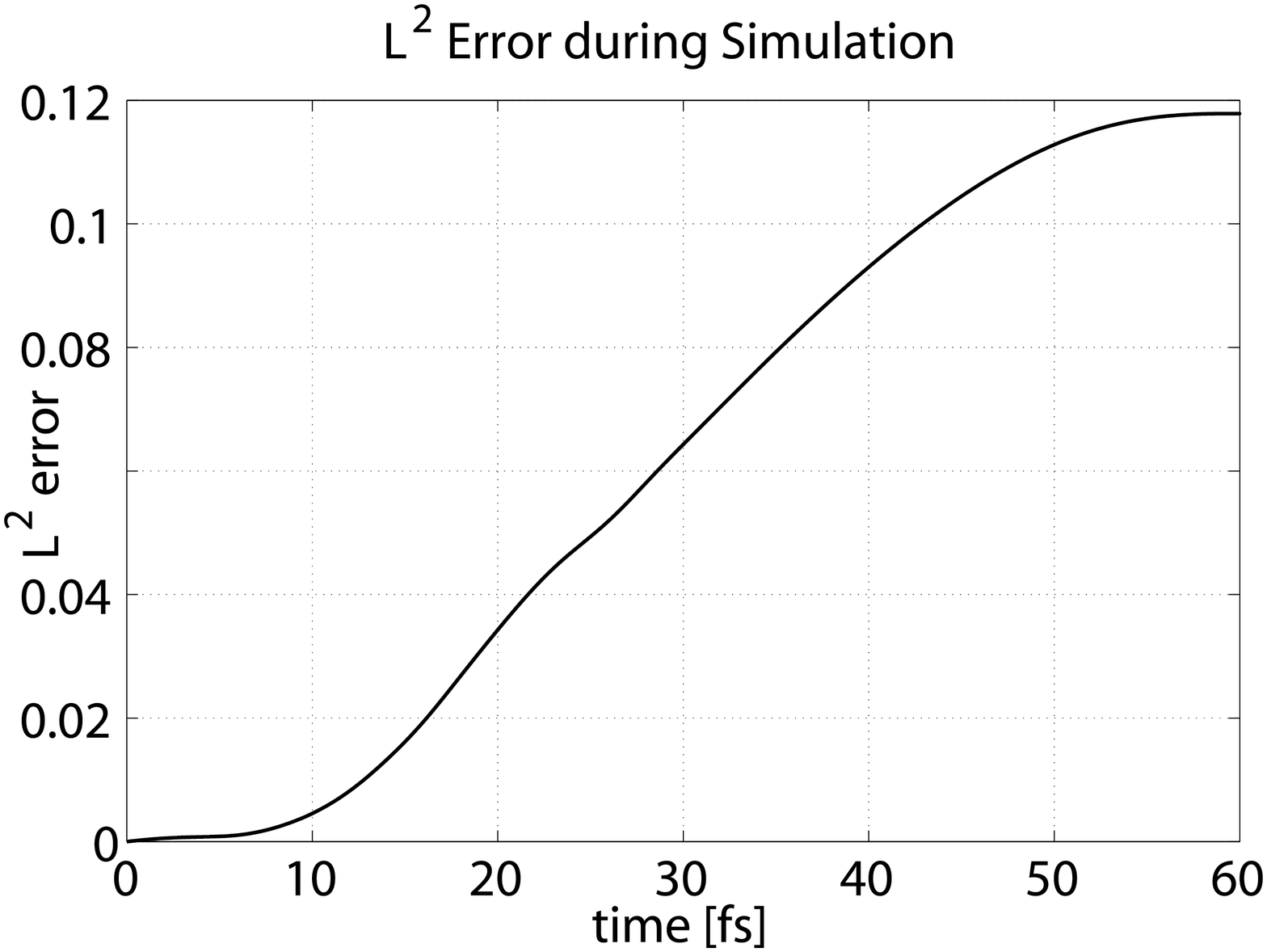} & \includegraphics[scale=0.24]{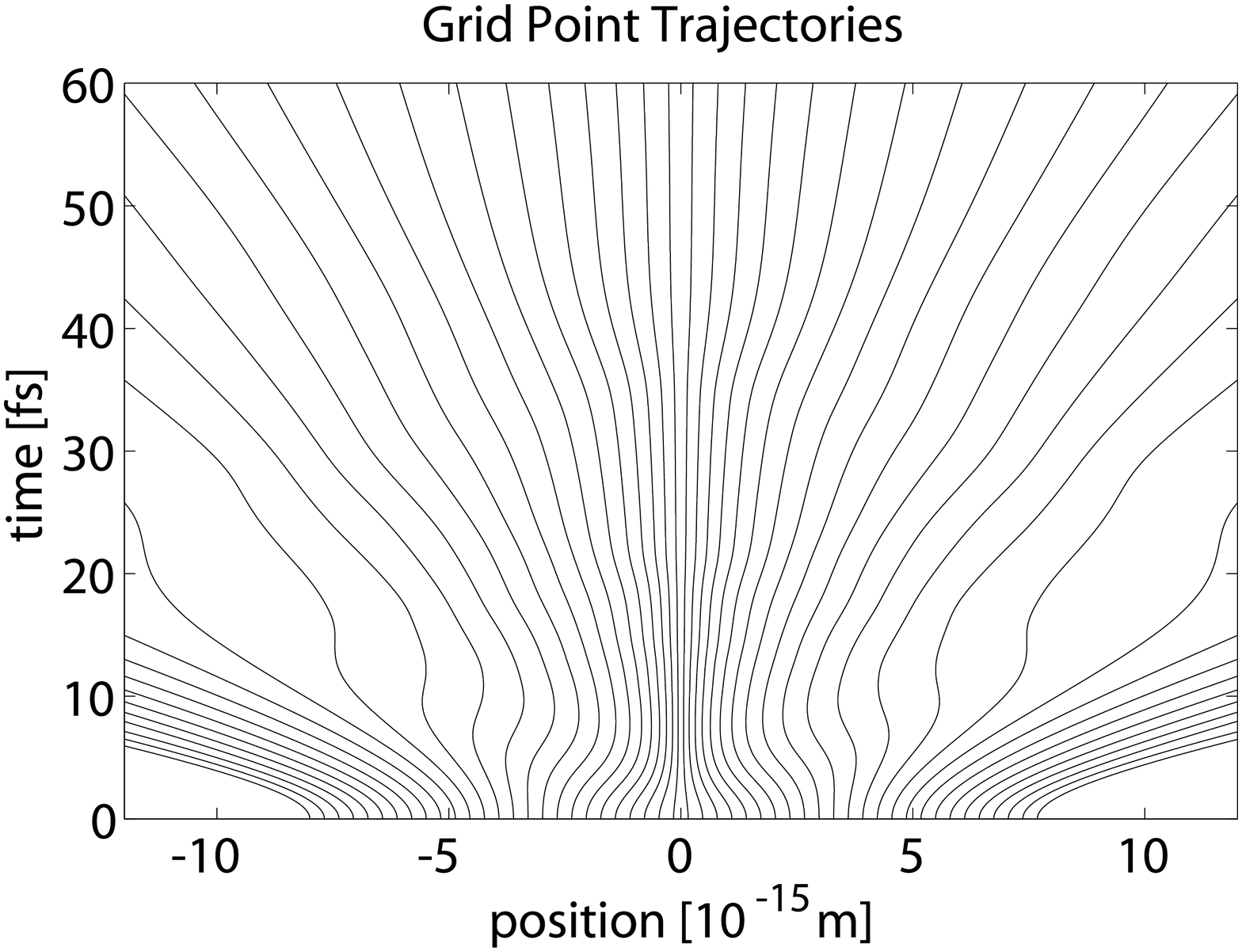}\\
\end{tabular}
\caption{Both plots belong to the simulation using polynomial fitting in between the boundary. Left: $L^2$ distance between the simulated wavefunction and the analytic solution of the Schr\"odinger equation. Right: A plot of the trajectories of the grid points. Note how some trajectories initially move towards each other, decelerate, and finally move apart.}
\label{fig:simulation_polyfit_error}
\end{figure}

\begin{figure}[hb]
\centering
\begin{tabular}{cc}
  \includegraphics[scale=0.24]{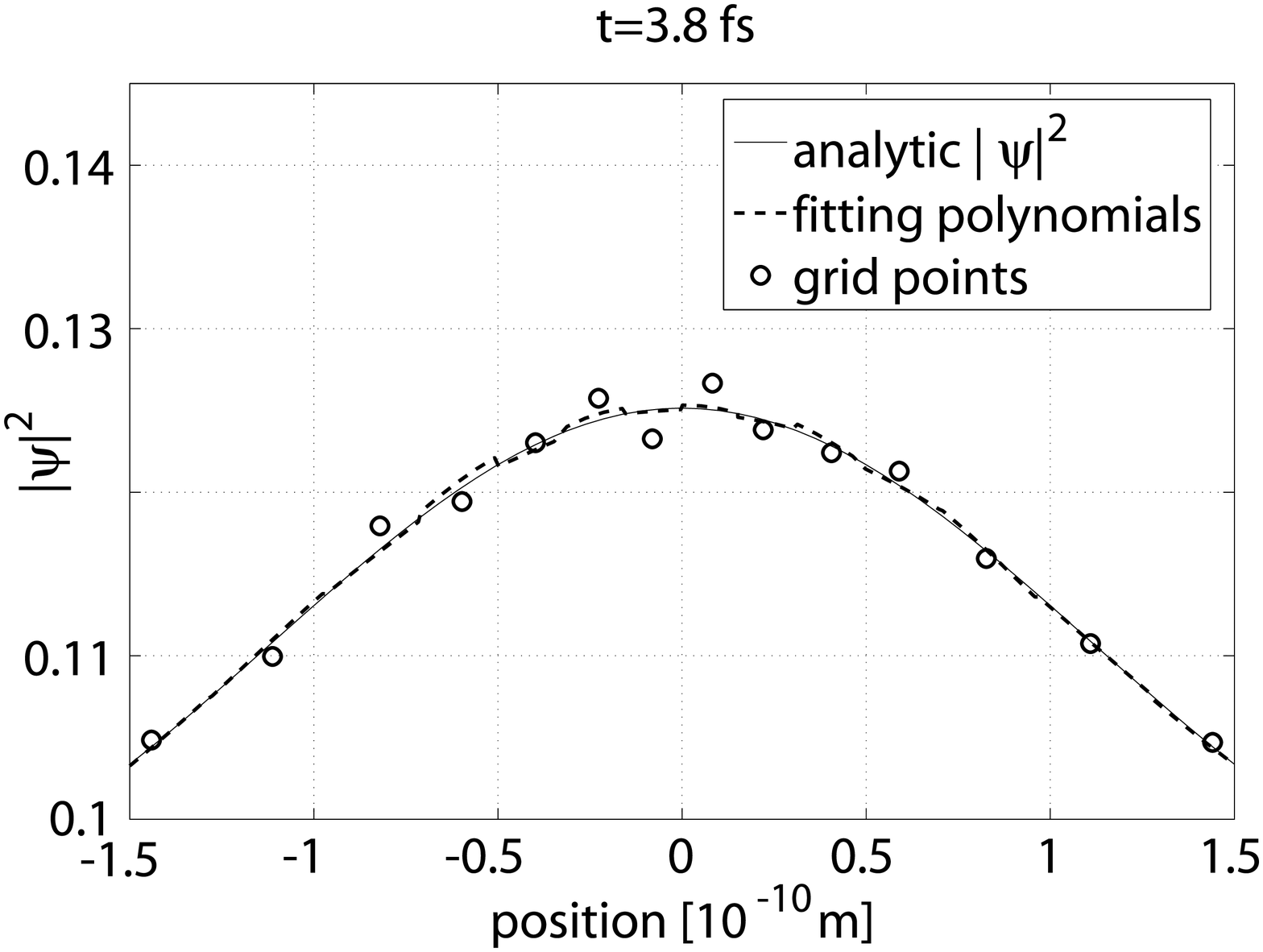} & \includegraphics[scale=0.24]{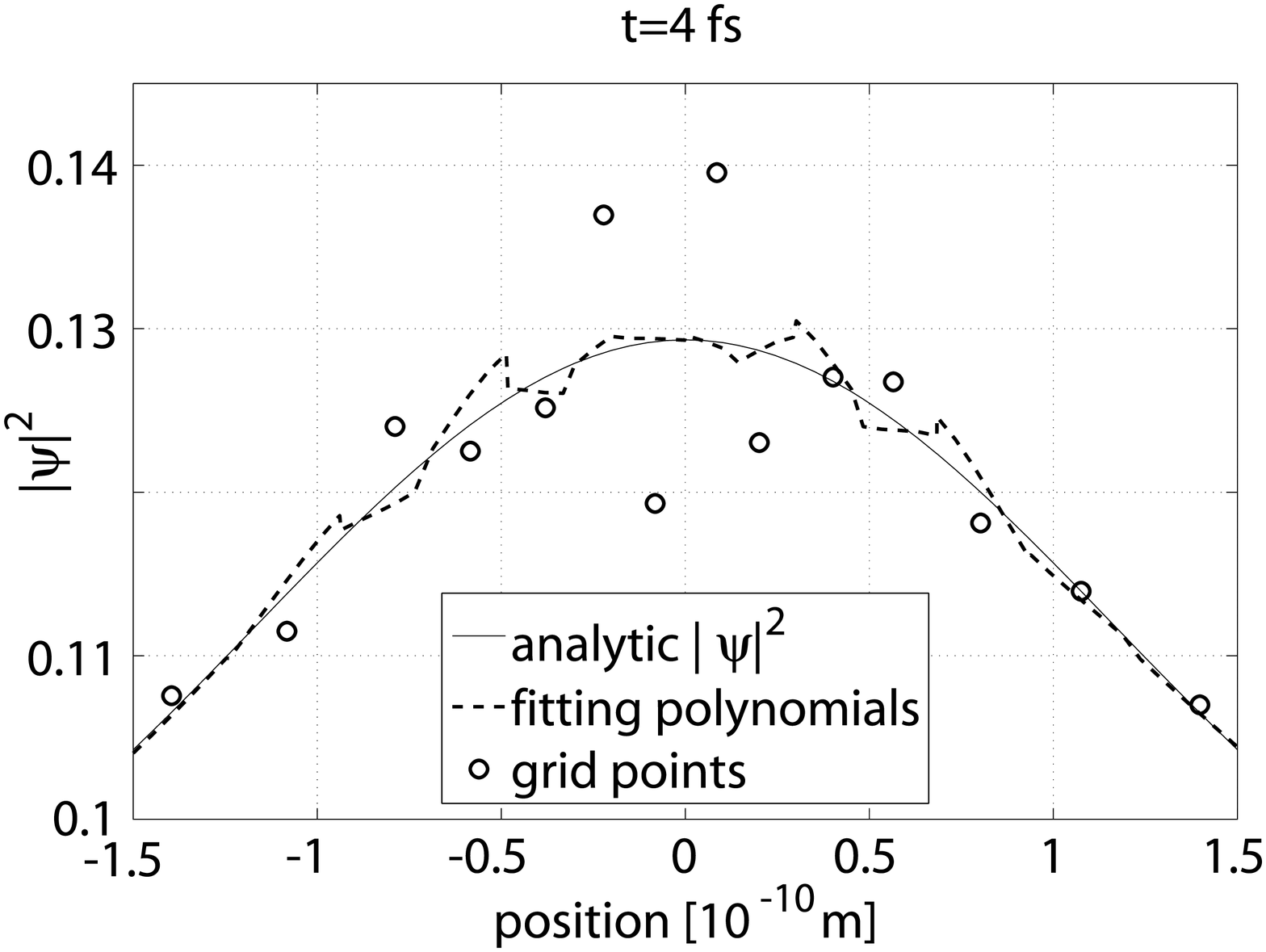}\\
  \includegraphics[scale=0.24]{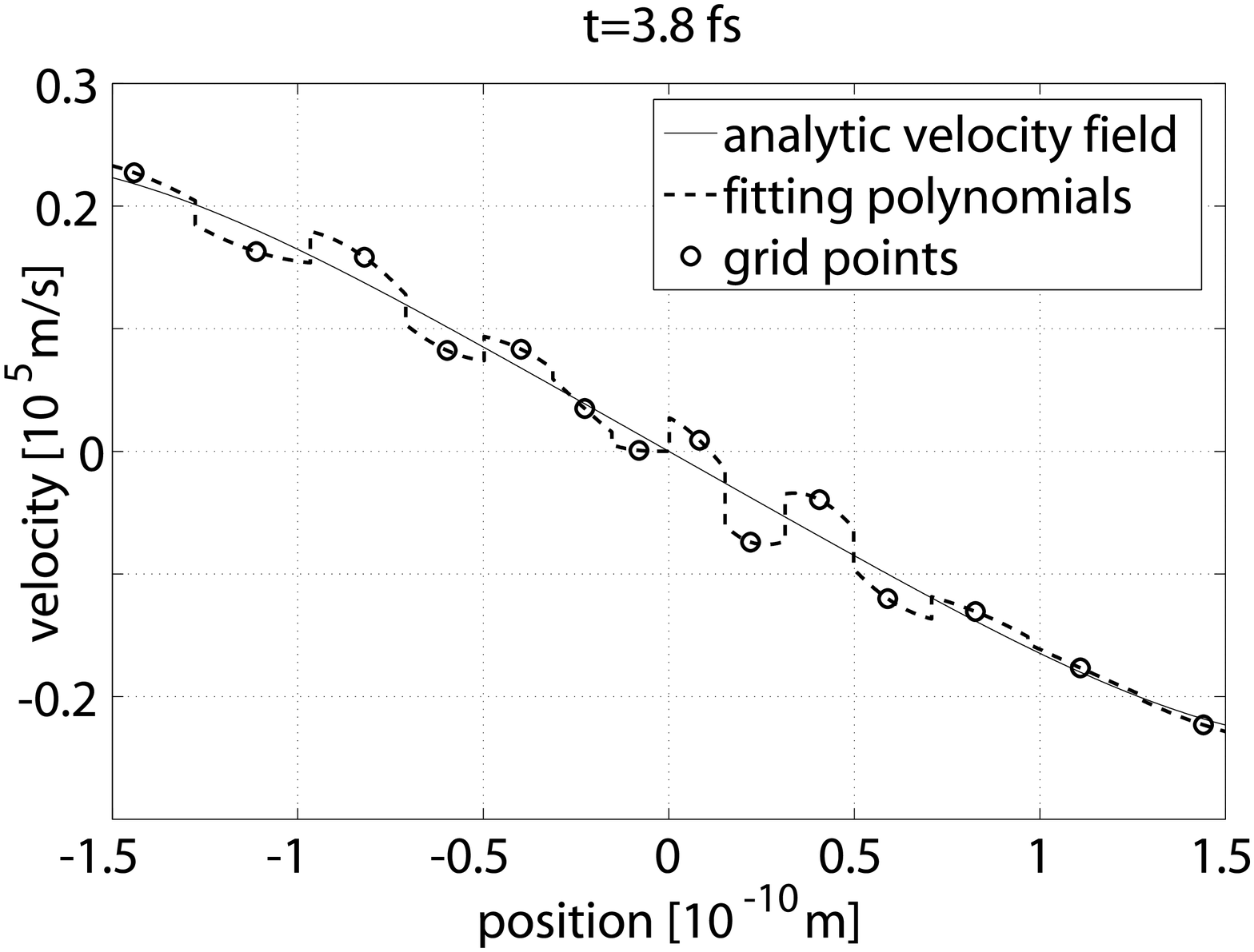} & \includegraphics[scale=0.24]{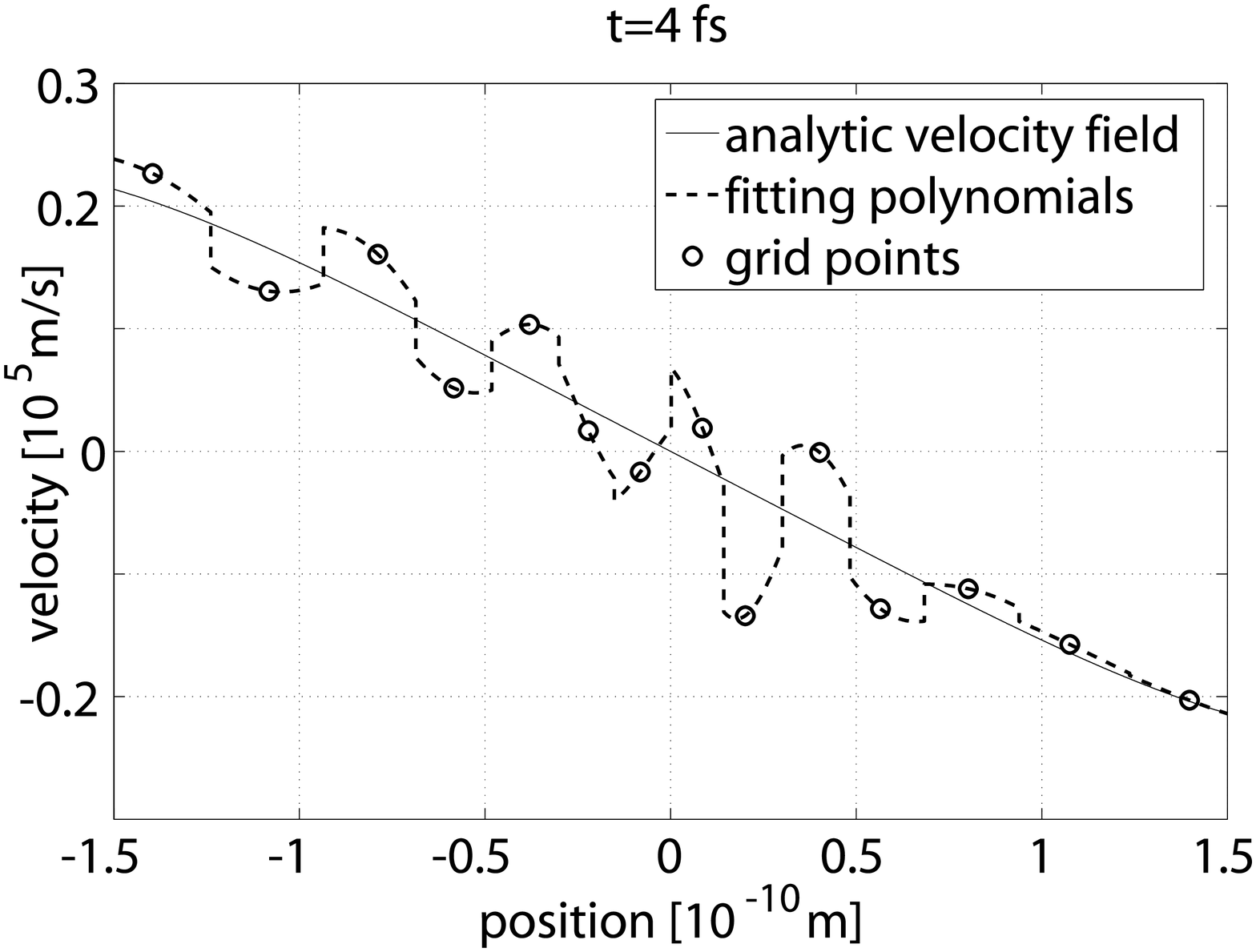}\\
\end{tabular}
\caption{Left: same as left-hand side of
\ref{fig:simulation_polyfit} but this time using least square
fitting throughout. Right: a short time later. Note in the upper
left plot, the fitting polynomials of least square fitting fail
to recognize the relatively big ordinate change of the grid
points. So the grid points moving towards each other are not
decelerated by the quantum potential. Finally after $t=4.3\text{fs}$ a
crossing of the trajectory occurs.}
\label{fig:simulation_leastsquare}
\end{figure}

\begin{figure}[hb]
\centering
\begin{tabular}{cc}
  \includegraphics[scale=0.24]{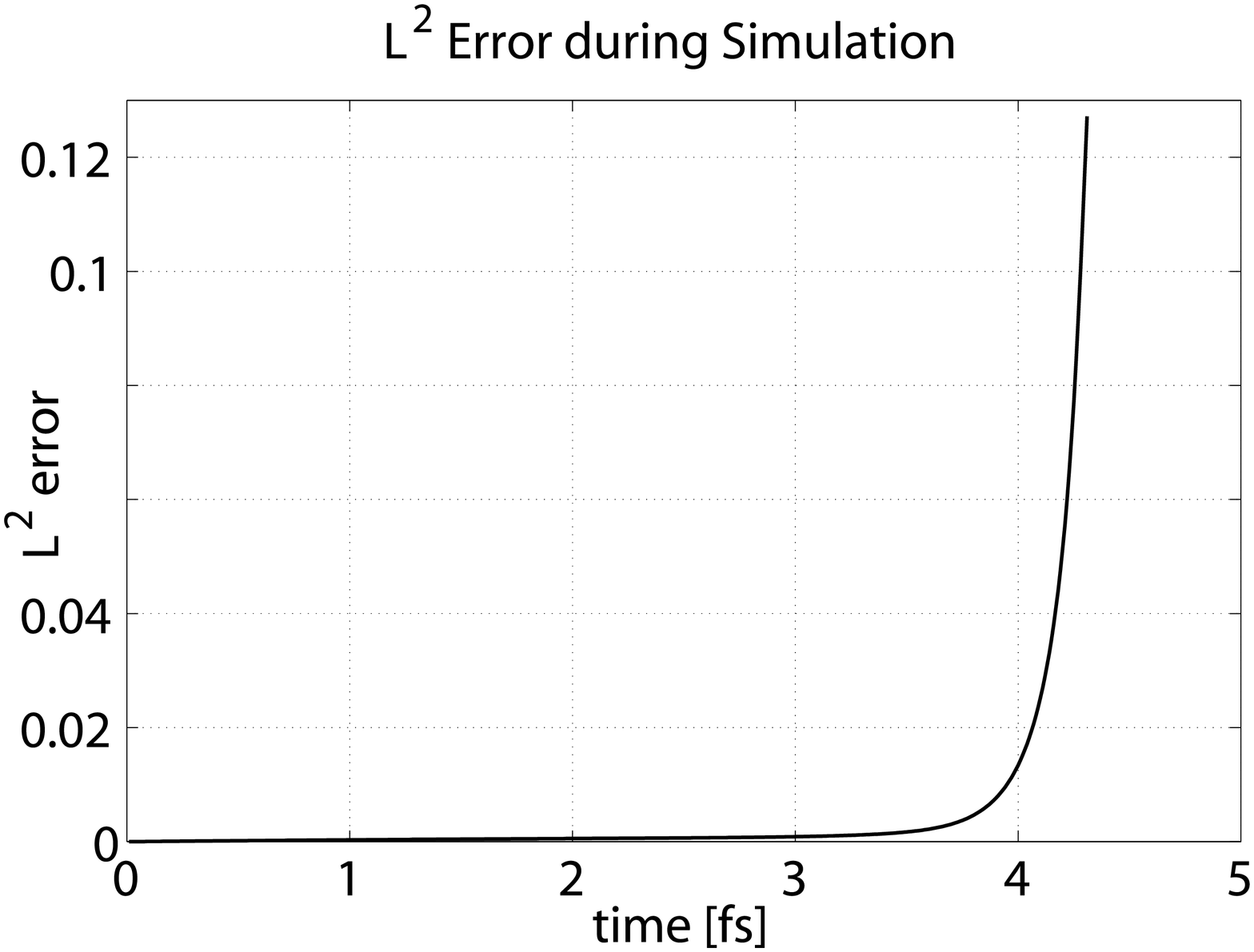} & \includegraphics[scale=0.24]{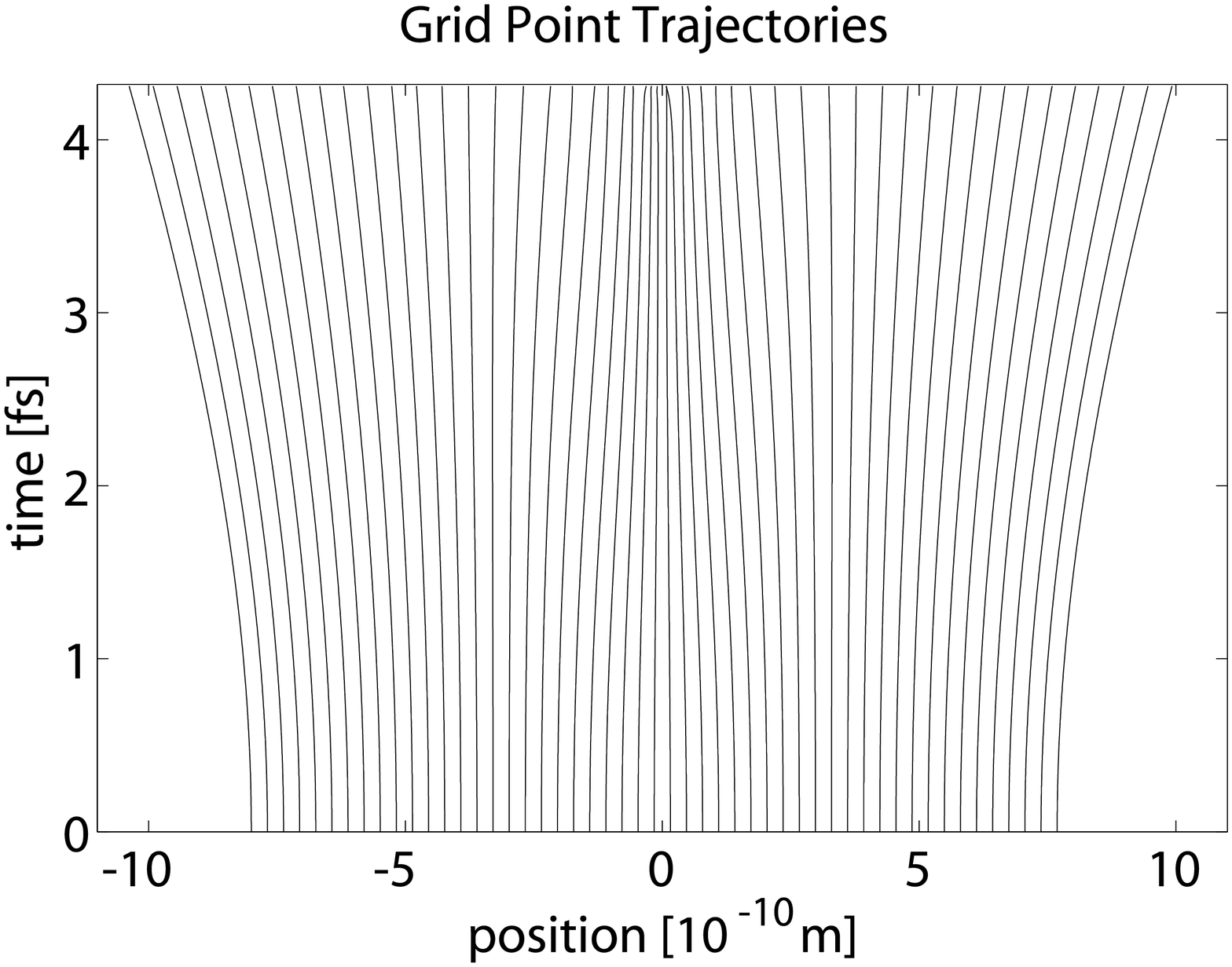}\\
\end{tabular}
\caption{Both plots belong to the simulation using least square fitting throughout. Left: $L^2$ distance between the simulated wavefunction and the analytic solution of the Schr\"odinger equation. Right: A plot of the trajectories of the grid points. Note by comparison with right-hand side of figure \ref{fig:simulation_polyfit_error} the failure of least square fitting to recognize grid point trajectories moving towards each other until they finally cross and the numerical simulation aborts.}
\label{fig:simulation_leastsquare_error}
\end{figure}

\paragraph{Results.} The first run with polynomial fitting yields
accurate results and does not allow for trajectory crossing way
beyond $5000$ integration steps, i.e. $50\text{fs}$, see figures \ref{fig:simulation_polyfit} and \ref{fig:simulation_polyfit_error}. The second run with least square fitting reports a crossing of trajectories
already after 430 integration steps, i.e. $4.3\text{fs}$, and aborts, see figures \ref{fig:simulation_leastsquare} and \ref{fig:simulation_leastsquare_error}. The numerical instability can already be observed earlier, see left-hand side of figure
\ref{fig:simulation_leastsquare}. An adjustment of the time step
of the numerical integration does not lead to better results in
the second run. The crossing of the trajectories of course occurs
exactly in the region were the grid points move fastest towards
each other. Referring to the discussion before, the small bumps in
$R$ created by the increase of the grid point density in this
region is not seen by the least square algorithm and thus not seen
by the numerical integration of equations
(\ref{eqn:diffEqs1})-(\ref{eqn:diffEqs3}), compare figure
\ref{fig:simulation_polyfit} and \ref{fig:simulation_leastsquare}.
The approaching grid points are not decelerated by the quantum
potential and finally cross each other, see figure
\ref{fig:simulation_leastsquare}, while in the first run they
begin to decelerate and turn to the opposite direction during the
time between $3\text{fs}$ to $5\text{fs}$, see figure
\ref{fig:simulation_polyfit}. To visualize how the two algorithms
"see" $R^2$ and the velocity field during numerical integration
these entities have been plotted in figures
\ref{fig:simulation_polyfit} and \ref{fig:simulation_leastsquare}
by merging all fitting polynomials in the neighborhood of every
grid point together (from halfway to the left neighboring grid
point to halfway to the right neighboring grid point). In figure
\ref{fig:simulation_leastsquare} one clearly observes the failure
of the least square fitting algorithm to see the small bumps in $R$.

Note that in some special situations in which the time step of the
numerical integration is chosen to be large, polynomial fitting
may lead to trajectory crossing as well. This may happen when the
number of time steps in which the simulation has to decelerate two fast
approaching grid points is not sufficient. This effect is entirely
due to the fact that numerical integration coarse grains the time.
The choice of a smaller time step for the simulation will always
cure the problem as long as other numerical errors do not
accumulate too much.


The relevant measure of quality of the numerical simulation is naturally the $L^2$ distance between the simulated wavefunction $Re^{iS}$ and the analytic solution of the Schr\"odinger equation $\psi_t$, i.e. $\left(\int dx\;|\psi_t(x)-R(x,t)e^{i S(x,t)}|^2\right)^{1/2}$, and is spelled out on the left-hand side of the figures \ref{fig:simulation_polyfit_error} and \ref{fig:simulation_leastsquare_error} for both runs.

\section{Using $|\psi^0|^2$ as Initial Distribution.}\label{dist}

As discussed in the introduction, if the grid point are distributed
according to $|\psi^0|^2$, they will remain so for all times. Doing
so will bring two advantages but comes at the price of a more severe
conceptual boundary problem as discussed in the last section. The
first advantage is that this Bohmian grid avoids regions where $R$
is very small and thus where the computation of the quantum
potential becomes numerically tricky. The second advantage is that
now the system is over-determined and the actual density of grid
points must coincide with $R^2$ for all times. This could be used as
an on the fly check whether the grid points behave like Bohmian
trajectories or not. If not the numerical integration does not give
a good approximation to the solution of
(\ref{eqn:diffEqs1})-(\ref{eqn:diffEqs3}). It could also be
considered to stabilize the numerical simulation with a feedback
mechanism balancing $R^2$ and the distribution of the particle
positions which may correct numerical errors of one or the other on
the fly.

One way of choosing $|\psi^0|^2$ distributed initial positions for
the grid points $q^0_j$ for $j=1\ldots n$ for some large number
$n$ is the following. Choose $q^0_j$ in such a way that
\begin{equation}\label{distribution}
    |\psi^0(q^0_j)|^2\frac{q^0_{j+1}-q^0_{j-1}}{2}=\frac{1}{n}
\end{equation}
This formula can be used to compute the grid points
iteratively\footnote{This procedure of setting the initial grid
points might run into a node of $\psi$. In that case one should
simply start with a slightly shifted initial grid point} starting
with some grid point near the maximum of $|\psi(q,0)|^2$. Another
way is to simulate $|\psi^0|^2$ distributed random variables $q^0_j$
by the commonly used techniques. Here, however, it has to be taken
care that the randomly chosen $q^0_j$ do not lie too close together.
If they do, either some of them must be deleted or the time step of
the numerical integration must be adjusted carefully. A rule of
thumb is that the distance of approaching grid points divided by
their relative velocity has to be a lot smaller than the chosen time
step.

\section{Conclusion}

Using polynomial fitting instead of least square fitting in between the boundary  increases the stability of the numerical integration immensely. This is due to the fact that polynomial fitting in contrary to least square fitting does not average over the microscopic structure of the function to fit and therefore reconstructs needed derivatives more accurately. The discussed boundary problem arises more severely for polynomial fitting because of this fact. However, this problem is generic to the numerical integration considered here and also occurs using least square fitting but in a milder way. We suggest that the smoothness of the decay of the wave function near the boundary should be the guide for solving this problem which requires a detailed study. Furthermore, we have discussed that the Bohmian grid is best adapted to the problem of numerical integration of (\ref{eqn:diffEqs1})-(\ref{eqn:diffEqs3}) because the grid points naturally avoid regions where $R$ becomes very small.

\section{Acknowledgment}
We thank Bob Wyatt for acquainting us with the Bohmian grid
methods in theoretical chemistry. We further thank in alphabetical
order: Lara Hernando, Salvador Miret-Artes, Angel Sanz, Clemens
Woywod for discussions. This work was partly supported by DFG.
D.-A. Deckert would like to cordially thank R.-R. Deckert and W.E.
G\"ohde-Deckert for funding and support.


\bibliographystyle{plain}

\newpage
\begin{verbatim}
SOURCE CODE OF THE NUMERICAL SIMULATION:

%% definition of constants and rendering of initial value

%CONSTANTS & SYMPBOLS*****************************************************

fovx            = 8;        %field of view on the real line [-fovx, fovx].
numParticles    = 51;       %number of Bohmian particles.
dt              = 10^-2;    %time step for numerical integration.
numSteps        = 6000;     %number of iterations of the numerical
                            %integration.
stencil         = 3;        %number of points left and right w.r.t. some
                            %point for fitting 0.5*log(rho).
degree          = 6;        %degree of polynomial for fitting
                            %please note: stencil=degree/2 means a
                            %precise polynomial fit of 'degree' degree
                            %while stencil>degree/2 means a least square
                            %fit with a polynomial of degree 'degree'.
s_stencil       = 3;        %log(sqrt(rho)) number of points left and
                            %right w.r.t. some point for fitting the
                            %phase.
s_degree        = 6;        %degree of polynomial for fitting the phase.
                            %the above note hols for the phase.

sym_x  = sym('sym_x','real');   %dummy symbol for symbolic math

%FUNCTIONS****************************************************************

%-------------------------------------------------------------------------
% Analytic solution of an initial gaussian evolved according to the free
% Schroedinger equation.
% INPUT : t=time of evolution, x=position, sigma=initial width of the
%         gaussian
% OUTPUT: complex value of the wavefunction
gauss  = @(t, x, sigma) (sigma/(pi*(sigma+i*t)^2))...
                        .^(1/4).*exp(-x.^2/(2*(sigma^2+t^2))*(sigma-i*t));

%-------------------------------------------------------------------------
% psi=analytic solution of a superposition of initial gaussians evolved
% according to the free Schroedinger equation.
% rho,S,vel=|psi|^2, the phase, the velocity field of psi
% INPUT : t=time of evolution, x=position
% OUTPUT: real resp. complex function value
psi    = @(t, x) 1/sqrt(2)*(gauss(t,x-3,4)+gauss(t,x+3,4));
rho    = @(t, x) psi(t,x).*conj(psi(t,x));
S      = @(t, x) imag(log(psi(t,x)./abs(psi(t,x))));
vel    = @(t, x) imag(subs(diff(psi(t,sym_x),sym_x)...
                 ./psi(t,sym_x),sym_x,x));

%INITIAL DATA-------------------------------------------------------------

clear x;                                %initial distribution of the
                                        %Bohmian particles
x = [-fovx:(2*fovx/numParticles):fovx]; %generate a uniform distribution

%plot the initial data: blue=initial rho and green=initial veleocity field
plot(x, vel(0,x), 'g', x, rho(0,x), 'b.-');
axis([-fovx fovx -1 1]);
grid on;

%% bohmian propagation

warning off;

%initialize the array that keeps track of the (x,0.5*log(rho),S,time)
%data history.
history = [{x vel(0,x) 0.5*log(rho(0,x)) S(0,x) 0}];

%loop over the numer of time steps of numerical integration
for (step = 1:numSteps)
    
    xlen = length(history{step,1});
    
    x_list          = history{step,1};
    vel_list        = history{step,2};
    logsqrtRho_list = history{step,3};
    s_list          = history{step,4};
    
    %first loop over the all Bohmian particles updating the
    %x_list and s_list
    for (c = 1:xlen)
        %find stencil points for logsqrtRho
        left_point  = c - stencil;
        right_point = c + stencil;
        deg = degree;
        if (left_point < 1)
            left_point  = 1;
            right_point = stencil*2+1 + round(numParticles/7);
            %decrease degree at the boundary as dodgy fix to avoid
            %bouadary problems
            deg = 2;
        elseif (right_point > xlen)
            left_point  = xlen - (2*stencil+1) - round(numParticles/7);
            right_point = xlen;
            %decrease degree at the boundary as dodgy fix to avoid
            %bouadary problems
            deg = 2;
        end
        %fit logsqrtRho_list
        fitdata_x = history{step,1}(left_point:right_point);
        fitdata_y = history{step,3}(left_point:right_point);
        logsqrtRho_fit = polyfit(fitdata_x, fitdata_y, deg);
        %compute derivatives
        d_logsqrtRho_fit = polyder(logsqrtRho_fit);
        dd_logsqrtRho_fit = polyder(d_logsqrtRho_fit);
        %compute quantum potential
        Q = -1/2*(polyval(dd_logsqrtRho_fit,x_list(c)) + ...
            polyval(d_logsqrtRho_fit,x_list(c)).^2);
        %update phase: S(t+dt)=S(t)+dt(1/2v(t)^2-Q(t))
        s_list(c) = s_list(c) + (1/2*vel_list(c).^2 - Q)*dt;
        %update position: r(t+dt)=r(t)+v(t)*dt
        x_list(c) = x_list(c) + vel_list(c)*dt;
    end
    
    %second loop over the all Bohmian particles updating
    %the vel_list and logsqrtRho_list
    for (c = 1:xlen)
        %find stecil points for S
        s_left_point  = c-s_stencil;
        s_right_point = c+s_stencil;
        deg = s_degree;
        if (s_left_point < 1)
            s_left_point  = 1;
            s_right_point = s_stencil*2+1 + round(numParticles/7);
            %decrease degree at the boundary as dodgy fix to avoid
            %bouadary problems
            deg = 2;
        elseif (s_right_point > xlen)
            s_left_point  = xlen - (2*s_stencil+1) - round(numParticles/7);
            s_right_point = xlen;
            %decrease degree at the boundary as dodgy fix to avoid
            %bouadary problems
            deg = 2;
        end
        %fit s_list
        fitdata_x = x_list(s_left_point:s_right_point);
        fitdata_y = s_list(s_left_point:s_right_point);
        s_fit = polyfit(fitdata_x, fitdata_y, deg);
        %compute derivatives
        d_s_fit = polyder(s_fit);
        dd_s_fit = polyder(d_s_fit);
        %v(t+dt) = S'(t+dt)
        vel_list(c) = polyval(d_s_fit,x_list(c));
        %R(t+dt) = R(t)*exp(-1/2 S''(t+dt)*dt)
        logsqrtRho_list(c) = logsqrtRho_list(c)-...
                             1/2*polyval(dd_s_fit,x_list(c))*dt;
    end
    
    %compile the tuple (x,vel,0.5*log(rho),time)
    newSnapshot = [{x_list vel_list ...
                    logsqrtRho_list s_list history{step,5}+dt}];
    %and save the snapshot
    history = [history; newSnapshot];
    
    %compute minimal distance of the Bohmian particles
    dxmin  = min(x_list(2:length(x_list))-x_list(1:length(x_list)-1));
    if (dxmin <= 0)
        sprintf('TRAJECTORY CROSSING OCCURED AT TIME: %f',...
                history{step,5}+dt)
        break;
    end
    
    if (step == 1 | mod(step/numSteps*100,10) == 0)
        %plot integrated velocity field
        plot(x_list(1):0.1:x_list(xlen),vel(history{step,5}+dt,...
             x_list(1):0.1:x_list(xlen)),'r',x_list,vel_list,'b.');
        axis([-25 25 -5 5]);
        grid on;
        sprintf('Progress: %.1f%%, timestep: %f fs, min deltax: %f',...
                step/numSteps*100, history{step,5}+dt, dxmin)
        pause(0.1);
    end
end

%% render rho movie (red is the analytic solution)
frame = 1;
clear('rhoMov');
for step=1:10:length(history)
    plot(-25:0.1:25,rho(history{step,5},-25:0.1:25),'r',...
         history{step,1},exp(history{step,3}*2),'b.',...
         history{step,1},0,'b*');
    axis([-25 25 0 0.3]);
    grid on;
    pause(0.05);
    rhoMov(frame) = getframe;
    frame = frame + 1;
end

%% render vel movie (red is the analytic solution)
frame = 1;
clear('velMov');
for step=1:10:length(history)
    plot(-20:0.1:20,vel(history{step,5},-20:0.1:20),'r',...
         history{step,1},history{step,2},'b.');
    axis([-25 25 -5 5]);
    grid on;
    velMov(frame) = getframe;
    frame = frame + 1;
end

%% render velocity field seen by the above algorithm

diagAxis = [-10 10 -1 1];

%loop over the number of time steps of numerical integration
for (step = 1:length(history))
    
    xlen = length(history{step,1});
    
    xvals = [];
    yvals = [];
    
    %loop over all particles
    for (c = 1:xlen)
        %find stecil points for S
        s_left_point  = c-s_stencil;
        s_right_point = c+s_stencil;
        deg = s_degree;
        if (s_left_point < 1)
            s_left_point  = 1;
            s_right_point = s_stencil*2+1 + round(numParticles/7);
            %decrease degree at the boundary as dodgy fix to avoid
            %bouadary problems
            deg = 2;
        elseif (s_right_point > xlen)
            s_left_point  = xlen - (2*s_stencil+1) - round(numParticles/7);
            s_right_point = xlen;
            %decrease degree at the boundary as dodgy fix to avoid
            %bouadary problems
            deg = 2;
        end
        %fit s_list
        fitdata_x = history{step,1}(s_left_point:s_right_point);
        fitdata_y = history{step,4}(s_left_point:s_right_point);
        s_fit = polyfit(fitdata_x, fitdata_y, deg);
        %compute velocity
        d_s_fit = polyder(s_fit);
        %compute graph
        
        if (c == 1)
            x1 = history{step,1}(c)-1;
            x2 = history{step,1}(c) ...
                 + (history{step,1}(c+1)-history{step,1}(c))/2;
        elseif (c == xlen)
            x1 = history{step,1}(c-1) ...
                 + (history{step,1}(c)-history{step,1}(c-1))/2;
            x2 = history{step,1}(c)+1;
        else
            x1 = history{step,1}(c-1) ...
                 + (history{step,1}(c)-history{step,1}(c-1))/2;
            x2 = history{step,1}(c) ...
                 + (history{step,1}(c+1)-history{step,1}(c))/2;
        end
            
        xvals = [ xvals x1:((x2-x1)/100):x2 ];
        yvals = [ yvals polyval(d_s_fit,x1:((x2-x1)/100):x2)];
    end
    
    plot(xvals,vel(history{step,5},xvals),'r',xvals,yvals,'b',...
         history{step,1},history{step,2},'b.');
    axis(diagAxis);
    grid on;
    pause(0.1);
end

%% render rho field seen by the above algorithm

diagAxis = [-30 30 0 0.145];

%loop over the number of time steps of numerical integration
for (step = 1:length(history))
    
    xlen = length(history{step,1});
    
    xvals = [];
    yvals = [];
    
    %loop over all particles
    for (c = 1:xlen)
        %find stencil points for logsqrtRho
        left_point  = c - stencil;
        right_point = c + stencil;
        deg = degree;
        if (left_point < 1)
            left_point  = 1;
            right_point = stencil*2+1 + round(numParticles/7);
            %decrease degree at the boundary as dodgy fix to avoid
            %bouadary problems
            deg = 2;
        elseif (right_point > xlen)
            left_point  = xlen - (2*stencil+1) - round(numParticles/7);
            right_point = xlen;
            %decrease degree at the boundary as dodgy fix to avoid
            %bouadary problems
            deg = 2;
        end
        %fit logsqrtRho_list
        fitdata_x = history{step,1}(left_point:right_point);
        fitdata_y = history{step,3}(left_point:right_point);
        logsqrtRho_fit = polyfit(fitdata_x, fitdata_y, deg);
        %compute derivatives
        d_logsqrtRho_fit = polyder(logsqrtRho_fit);
        dd_logsqrtRho_fit = polyder(d_logsqrtRho_fit);
        %compute quantum potential
        Q = -1/2*(polyval(dd_logsqrtRho_fit,x_list(c)) + ...
            polyval(d_logsqrtRho_fit,x_list(c)).^2);
        %compute graph
        
        if (c == 1)
            x1 = history{step,1}(c)-1;
            x2 = history{step,1}(c) ...
                 + (history{step,1}(c+1)-history{step,1}(c))/2;
        elseif (c == xlen)
            x1 = history{step,1}(c-1) ...
                 + (history{step,1}(c)-history{step,1}(c-1))/2;
            x2 = history{step,1}(c)+1;
        else
            x1 = history{step,1}(c-1) ...
                 + (history{step,1}(c)-history{step,1}(c-1))/2;
            x2 = history{step,1}(c) ...
                 + (history{step,1}(c+1)-history{step,1}(c))/2;
        end
            
        xvals = [ xvals x1:0.05:x2 ];
        yvals = [ yvals polyval(logsqrtRho_fit,x1:0.05:x2)];
    end
    
    plot(xvals,rho(history{step,5},xvals),'r',xvals,exp(2*yvals),'b',...
         history{step,1},exp(2*history{step,3}),'b.');
    axis(diagAxis);
    grid on;
    pause(0.1);
end

%% render trajectories

diagAxis = [-12 12 0 length(history)*dt];
hold off;
grid off;
        
%loop over the grid points
for (j = 1:numParticles)
    xvals = [];
    yvals = [];
    for (step = 1:length(history))
        xvals = [xvals; history{step,1}(j)];
        yvals = [yvals; history{step,5}];
    end

    plot(xvals,yvals,'k-');
    axis(diagAxis);
    hold on;
    pause(0.1);
end    
hold off;

%% render L^2 error
psi2error = [];
for (step = 1:length(history))
    xdiff  = [];
    ydiff2 = [];
    for (j = 1:numParticles-1)
        xdiff = [xdiff; history{step,1}(j+1)-history{step,1}(j)];
        ydiff2 = [ydiff2; abs(exp(complex(0,history{step,4}(j+1)))...
                              .*exp(history{step,3}(j+1))...
                              - psi((step-1)*dt,history{step,1}(j+1)))^2];
    end
    psi2error = [psi2error; sqrt(sum(xdiff.*ydiff2))];
end
plot((1:length(history))*dt,psi2error,'k');
\end{verbatim}

\end{document}